\documentclass[twocolumn,english,aps,PRA,reprint, superscriptaddress,showpacs,longbibliography,showkeys]{revtex4-1}
\usepackage{amsmath,amssymb,bbm,mathrsfs,bm,braket,color,graphicx,comment}
\usepackage[colorlinks,citecolor=blue,urlcolor=blue,linkcolor = blue]{hyperref}
\usepackage[mathscr]{euscript}

\newcommand\minus{
  \setbox0=\hbox{-}
  \vcenter{
    \hrule width\wd0 height \the\fontdimen8\textfont3
  }%
}

\makeatother
\newcommand{\abs}[1]{\left\lvert #1 \right\rvert}

\begin{document}

\title{Designing ground states of Hopfield networks for quantum state preparation}
\author{Clemens Dlaska}
\email{clemens.dlaska@uibk.ac.at}
\affiliation{Institute for Theoretical Physics, University of Innsbruck, A-6020 Innsbruck, Austria}
\author{Lukas M. Sieberer}
\email{lukas.sieberer@uibk.ac.at}
\affiliation{Center for Quantum Physics, Faculty of Mathematics, Computer Science and Physics, University of Innsbruck, 6020 Innsbruck, Austria}
\affiliation{Institute for Quantum Optics and Quantum Information of the Austrian Academy of Sciences, A-6020 Innsbruck, Austria}
\author{Wolfgang Lechner}
\email{w.lechner@uibk.ac.at}
\affiliation{Institute for Theoretical Physics, University of Innsbruck, A-6020 Innsbruck, Austria}
\affiliation{Institute for Quantum Optics and Quantum Information of the Austrian Academy of Sciences, A-6020 Innsbruck, Austria}
\email{w.lechner@uibk.ac.at}

\begin{abstract}
We present a protocol to store a polynomial number of arbitrary bit strings, encoded as spin configurations, in the approximately degenerate low-energy manifold of an all-to-all connected Ising spin glass. The iterative protocol is inspired by machine learning techniques utilizing $k$-local Hopfield networks trained with $k$-local Hebbian learning and unlearning. The trained Hamiltonian is the basis of a quantum state-preparation scheme to create quantum many-body superpositions with tunable squared amplitudes using resources available in near term experiments. We find that the number of configurations that can be stored in the ground states and thus turned into superposition scales with the $k$-locality of the Ising interaction. 

\end{abstract}
\pacs{}
\maketitle

\section{Introduction}
\label{sec:Intro}

Preparation and control of quantum many-body superpositions is a cornerstone of current efforts in quantum simulation and quantum computation~\cite{Cirac2012,Georgescu2014,DiCarlo2010,Bernien2017,Raimond2001}. In particular, quantum algorithms such as a quantum solver for linear systems of equations~\cite{HHL2009}, quantum support vector machines~\cite{Rebentrost2014}, quantum principal component analysis~\cite{Lloyd2014} and other quantum machine learning algorithms~\cite{Biamonte2017,Lloyd2013,Dunjko2018} rely on an input state that contains data as a quantum superposition. However, a universal device that transforms classical data into a quantum superposition requires exponential resources, and its implementation is considered one of the major challenges in quantum computing. The pioneering proposal for such a device is known as quantum random access memory (QRAM)~\cite{Giovanetti2008a,Giovanetti2008b}. The physical implementation of a general gate-based QRAM scheme requires coherent control over exponential resources in the length of bit strings. Thus, it is a natural question, whether one can find protocols that are less general, but in turn, less hardware-intensive,  which is particularly relevant for near term quantum devices~\cite{Preskill2018}. 

Recently, a scheme based on Hamiltonian quantum state preparation has been proposed \cite{Sieberer2018} to prepare a superposition of a polynomial number of bit strings with programmable squares of the amplitudes. The obtained states are phase coherent, but the individual phases are not programed. In this scheme, it is assumed that it is possible to encode a polynomial number of bit strings as $M$-fold degenerate ground states of an Ising spin system. The desired quantum many body superposition is then prepared in an adiabatic-diabatic protocol which transforms a trivial product state into a superposition of the $M$ spin configurations. The method in Ref. \cite{Sieberer2018} requires a polynomial number of qubits, and the realization of a particular final Hamiltonian $H_f$ may require all possible $k$-body Ising spin interactions $(k=1,\dots,N)$.

Here, we present a variational protocol to design Ising Hamiltonians with approximately degenerate ground states composed of a polynomial number of $M$ configurations utilizing resources available in near term experiments [i.e. pair interactions $k=2$ and three-body interactions $k=3$ implemented with $O(N^k)$ qubits]. Inspired by machine learning techniques we use a $k$-local Hopfield network ~\cite{Hopfield1982, Hebb1949, Baldi1987} as an ansatz to design the energy spectrum. This ansatz is then variationally optimized via an iterative $k$-local Hebbian relearning and unlearning protocol.
\begin{figure}[t]
\begin{centering}
\includegraphics[width=\columnwidth]{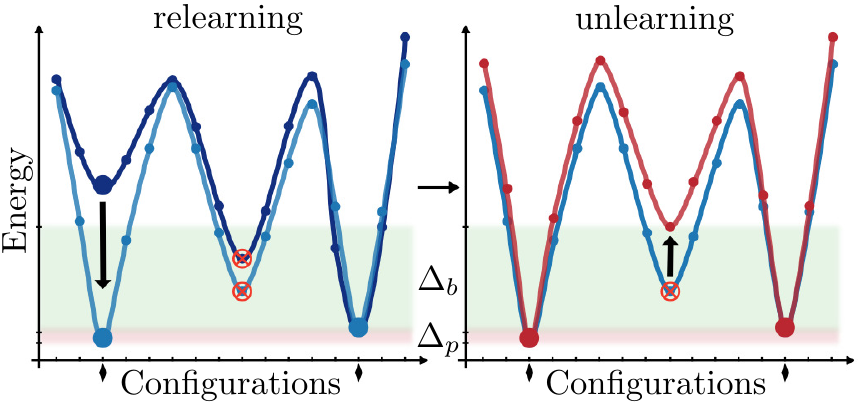}
\par\end{centering}
\protect\caption{Schematic of the ground-state design method based on Hopfield networks. (Left panel) Initially, the patterns (indicated by diamonds) are stored in a Hopfield network. Thus, patterns are located at local energy minima [dark blue (upper) curve]. Relearning of a single pattern [transition from dark blue (upper) to light blue (lower) curve] decreases the energy bandwidth $\Delta_p$ of the stored patterns. (Right panel) Unlearning of individual low-energy bulk configurations [transition from light blue (lower) to red (upper) curve], which are typically spurious minima (indicated by crossed circles), increases the energy gap $\Delta_b$ between stored patterns and the bulk. Iteratively applying relearning and unlearning steps results in (approximately) degenerate ground states with $\Delta_p/\Delta_b\ll 1$.}
\label{fig0}
\end{figure}
The reasons for utilizing Hopfield networks as an ansatz are twofold. First, storing patterns in energy minima of an Ising spin-glass Hamiltonian strongly resembles the notion of learning patterns in Hopfield networks. Second, Hopfield networks are based on low $k$-local terms, which is in contrast to $N$-local Ising interactions needed for an exact expansion of a particular final Hamiltonian in terms of Ising interactions. As an example consider $H_f$ to be the projector onto the data bit strings $\ket{x_n}=\ket{1011100\dots}$~\footnote{The states $\ket{x_n}=\ket{1011100\dots}$ are regarded as product states in the Pauli $\sigma_z$ basis, with individual bits $x_{n,i} = 0,1$ corresponding to eigenvalues $\pm1$ of $\sigma_z^{(i)}$} of the form
\begin{equation}
\label{eq:projector}
H_f = \mathbbm{1}-\sum_{n=1}^M\ket{x_n}\bra{x_n}.
\end{equation}
Expanding Eq.~\eqref{eq:projector} in terms of individual $\sigma_z^{(i)}$ Pauli operators results, apart from a global energy offset, in an all-to-all connected $N$-local Ising spin Hamiltonian of the form
\begin{eqnarray}
\label{eq:decomposition}
H_f &=&\sum_{i} J_i\sigma_z^{(i)}+\sum_{i<j}J_{ij}\sigma_z^{(i)} \sigma_z^{(j)}\nonumber\\
& & +\sum_{i<j<k}J_{ijk}\sigma_z^{(i)}\sigma_z^{(j)}\sigma_z^{(k)} + \dots,
\end{eqnarray}
where the number of necessary parameters represented by the number of matrix elements $\{J_i,J_{ij},J_{ijk}, \dots \}$ scales as $O(2^N)$.

Our protocol is illustrated in Fig.~\ref{fig0}. The goal is to construct an energy spectrum with $\Delta_p$, the energy bandwidth of the stored patterns, small compared to the energy gap $\Delta_b$ which separates the stored patterns from the ($2^N-M$) bulk configurations. This is achieved via a two step process:

\paragraph{Initialization as a $k$-local Hopfield network:} A Hopfield network is constructed as an ansatz Hamiltonian where the interactions are determined from applying the Hebbian learning rule \cite{Hebb1949} on all $M$ configurations to be in the ground state. This guarantees that the configurations are local energy minima of the spectrum [dark blue (upper) curve in left panel of Fig.~\ref{fig0}].
\paragraph{Variational ground-state design:} The interaction matrix elements of the initial $k$-local Hopfield network are modified by applying Hebbian relearning or unlearning~\cite{Hopfield1983,Kleinfeld1987,Fachechi2018} steps on individual configurations. Due to the specific form of the ansatz Hamiltonian, these relearning [cf. left panel of Fig.~\ref{fig0}: dark blue (upper) to light blue (lower) curve] and unlearning steps [cf. right panel of Fig.~\ref{fig0}: light blue (lower) to red (upper) curve] allow one to dominantly shift individual configurations in energy down or up, respectively, without inducing major shifts in the bulk states of the all-to-all connected spin model. The second step of the protocol is iterated in a Monte Carlo fashion in order to variationally optimize the Hamiltonian towards approximate degeneracy ($\Delta_p/\Delta_b\ll 1$).

We find that the approximate Hamiltonians after the variational optimization can store, at least, $O(N^{k-1})$ patterns with almost exact degeneracy. We demonstrate the applicability of our approximate Hamiltonian for the protocol in Ref.~\cite{Sieberer2018} and extend the framework and effective theory to nondegenerate ground states and to three-local target Hamiltonians.

The remainder of this paper is organized as follows. First, in
Sec.~\ref{sec:generalhopf}, we give a short review of Hofpield networks including
a general $k$-local version of the Hopfield network with $k$-local Hebbian learning and the resulting theoretical storage capacity. Based on this, we describe in detail our iterative Monte Carlo ground-states design protocol in Sec.~\ref{sec:GSdesign}. Furthermore, we discuss the capacity of our approach in Sec.~\ref{sec:capacity}. In Sec.~\ref{sec:PS}, we extend the framework proposed in Ref.~\cite{Sieberer2018} to generate programmable superpositions of many-body states by allowing for spin models with higher-order interactions where the data bit strings can also be encoded in nondegenerate low-energy states. At the end of this section we show examples of our full approach in two-dimensional (2D) and three-dimensional (3D) Lechner-Hauke-Zoller (LHZ)~\cite{Lechner2015} architectures. In Sec.~\ref{sec:conclusion}, we conclude and give an outlook on future research directions.

\section{Ground-state design}
\subsection{$k$-local Hopfield network with a $k$-local Hebbian learning rule}
\label{sec:generalhopf}

\begin{figure*}[t]
  \begin{centering}
  \includegraphics[width=0.9\textwidth]{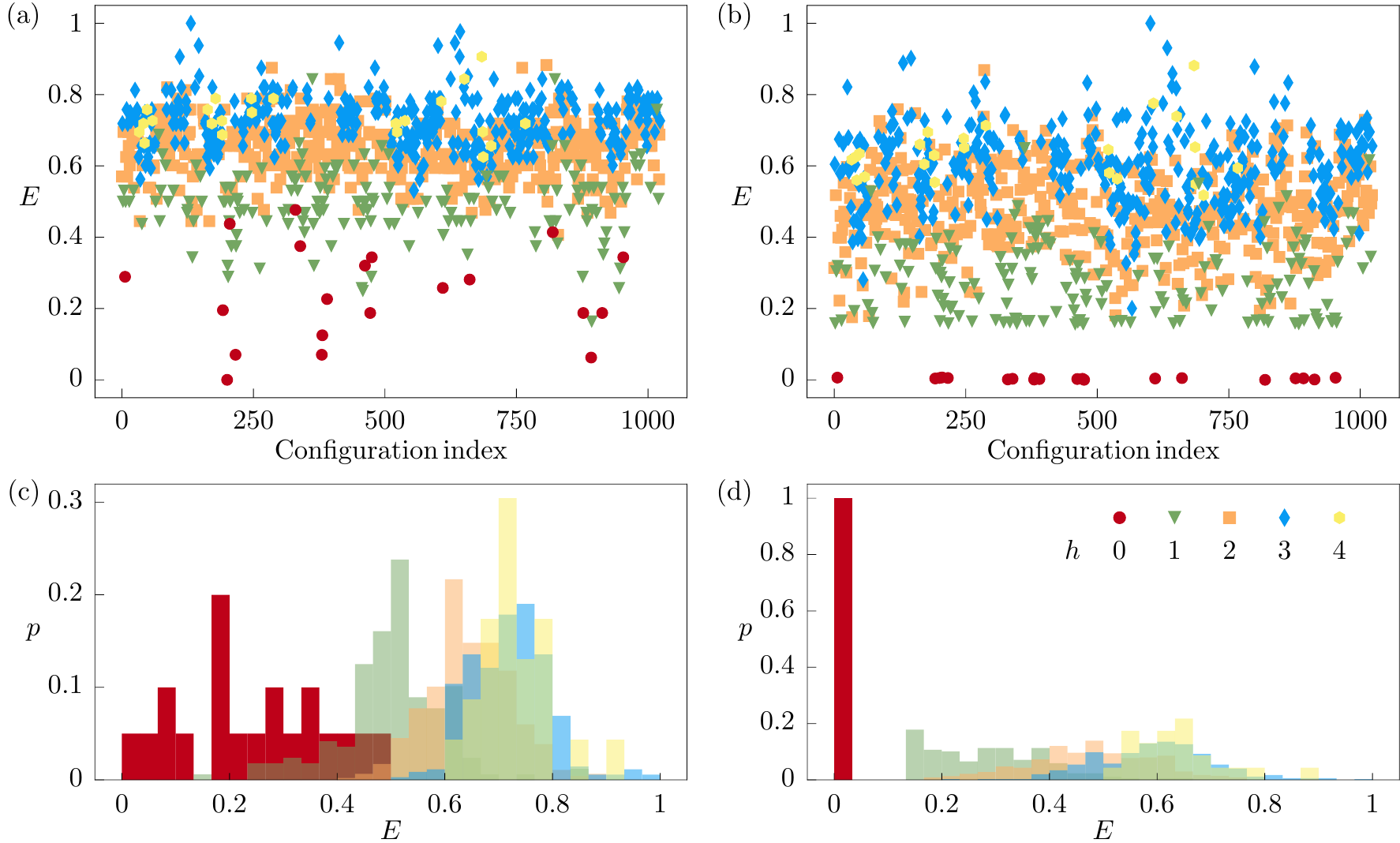}
  \par\end{centering}
\protect\caption{Typical spectra for a ground-state design of a system with $N = 10$, $K = \{2,3\}$, and $M = 20$ randomly chosen patterns. The minimum Hamming distance $h$ of every configuration with respect to any stored pattern is color (symbol) coded. (a) Spectrum of the initial Hopfield network trained with all $M$ patterns. (b) Spectrum after the iterative ground-state design protocol with $\Delta^* = 0.05$. The distribution of energies with respect to the Hamming distance $h$ (dark gray: $h=0$; light gray: $h\geq1$) is shown for the initial step of the protocol (c) and after the protocol (d). For this example, we use $h=2$, and thus 654 configurations out of $1024$ states. To achieve  $\Delta = 0.044$, $t_f=1820$ iteration steps were needed. Further parameters are $p_\mathrm{relearn}=2/3$, $p_\mathrm{unlearn}=1/3$, $0\leq\phi_k,\eta_k\leq 0.02$, $r = 1$, and $T=1$.}
\label{fig1}
\end{figure*}

The original Hopfield network~\cite{Hopfield1982} is a fully connected two-local graph with $N$ nodes (or neurons), which can be written as a spin-glass Hamiltonian of the form
\begin{equation}
H_{\mathrm{hf}} = \sum_i\theta_i\sigma_z^{(i)} + \sum_{i<j}J_{ij}\sigma_z^{(i)} \sigma_z^{(j)}.
\label{hf}
\end{equation}
This network is characterized by the interaction matrix elements $J_{ij}$ and local field terms $\theta_i$. The interaction matrix $J_{ij}$ of the network is constructed by suitable learning rules. The most prominent learning rule is the Hebbian learning rule~\cite{Hebb1949},
\begin{equation}
J_{ij} = -\frac{1}{M}\sum_{m=1}^Mx_i^mx_j^m,
\end{equation}
where $M$ is the number of bit strings to be stored and $x_i^m\in\{+1,-1\}$ is the eigenvalue of $\sigma_z^{(i)}$ of bit-string $m$. The Hebb rule aims at storing $M$ patterns as local minima of the energy spectrum. Configurations stored as local minima are also called stable states of a neuronal network and fulfill the stability condition,
\begin{equation}
\label{eq:thresholdfunction}
\mathrm{sgn}\left(\sum_{j\neq i}J_{ij}x_j^m+\theta_i\right) = x_i^m,\qquad\forall i.
\end{equation}
Note that Eq.~\eqref{eq:thresholdfunction} can be interpreted as a linear threshold function for neuron $i$. If Eq.~\eqref{eq:thresholdfunction} is fulfilled, the state is a local energy minimum with respect to the Hamming distance. Using these threshold functions, one can show that the maximum number of patterns a two-local Hopfield net can store as stable states of the network is at most equal to the number $N$ of available neurons \cite{Mostafa1985}. 

However, the storage capacity can be increased by allowing for higher-order $k$-local interactions~\cite{Baldi1987,Baldi1988}.
In general, including $k$-body interactions with $k\in K\subseteq\{1,\dots,N\}$, the Hamiltonian of the spin model can be written as
\begin{equation}
\label{eq:Hfull}
H = \sum_{k \in K}\sum_{\chi\in I}J_\chi\prod_{i=1}^k\sigma_z^{(\chi_i)},
\end{equation}
where $I = \lbrace (\xi_1,\cdots,\xi_k)\,\vert\,\xi_i \in \{1, \dotsc N\} \text{ and } \xi_1<\xi_2<\dots<\xi_k\rbrace$ is the set of indices labeling particular interactions between neurons. The interaction matrix elements $J_\chi$ can be obtained via the $k$-body Hebbian learning rule,
\begin{equation}
\label{eq:klearning}
J_\chi =- \frac{1}{M}\sum_{m = 1}^M\prod_{i=1}^kx_{\chi_i}^m.
\end{equation}

Including higher-order interactions increases the storage capacity of the network due to the increase of parameters available. Furthermore, higher-order correlations between patterns can be resolved, which are ``invisible'' for two-body interactions. 

The upper bound on the information storage capacity of $k$-local Hopfield networks can be estimated using threshold logic arguments similar to those first developed in the context of two-local Hopfield networks~\cite{Mostafa1985}. The upper bound for $M$ arbitrary patterns to be stable in a $k$-local Hopfield network is proportional to the number of parameters defined by the available interaction matrix elements. Thus, for a Hopfield net with single- to $d$-body interactions, the upper bound for the storage capacity is given by (for details see Ref.~\cite{Baldi1987,Baldi1988})
\begin{equation}
\label{eq:capacity}
M\leq\sum_{i=1}^{d-1}\binom{N-1}{i}.
\end{equation}
This upper bound is quite general and can be refined to more detailed bounds for particular learning rules. Nevertheless, all of them have in common that the maximum storage capacity is of order $O(N^{d-1})$.
Thus, for the experimentally realistic case of $K = \{1,2,3\}$ one can store at most $M \propto O(N^2)$ arbitrary patterns as stable states of the system. Equation~\eqref{eq:capacity} also reproduces the maximum storage capacity of $M\approx 2^N$ for $d=N$. As shown in Fig.~\ref{fig1}(a), the learned patterns are local energy minima of the spectrum given by an example Hamiltonian Eq.~\eqref{eq:Hfull}.

 In the conventional usage of classical Hopfield networks, the network is fixed after the training phase and serves as content addressable memory (CAM)~\cite{Hertz1991} in the subsequent recall phase. Also, several quantum-mechanical generalizations of Hopfield networks serving as quantum CAMs have been proposed~\cite{Neigovzen2009,Santra2017,Fard2018,Rebentrost2018,Seddiqi2018,Rotondo2018}. We note, that compared to Ref. \cite{Neigovzen2009} our scheme can be implemented with a low k-body Ising-type Hamiltonian.  
  
 Here, in contrast, we want to use the variationally trained classical Hopfield network as starting point for an output mechanism, based on quantum annealing, which aims at providing a controllable quantum superposition state composed of all learned patterns. As a prerequisite for this type of ``quantum recall'', the patterns need to be not just local minima but rather nearly degenerate ground states of the classical Ising spin system.

\subsection{Ground-state design protocol}
\label{sec:GSdesign}

In this section, we present the details of our variational ground-state design method with the goal to achieve a situation in which the energy bandwidth $\Delta_p$ of the stored patterns is small compared to the energy gap $\Delta_b$ separating the patterns from the bulk states, i.e., $\Delta=\Delta_p/\Delta_b\ll 1$ (cf. Fig.~\ref{fig0}). 

As mentioned in Sec.~\ref{sec:Intro}, our protocol consists of two major steps.
\subsubsection{Initialization as a $k$-local Hopfield network}
As the starting point of our protocol, $M$ bit strings of length $N$, that we want to bring into superposition, are encoded as spin configurations and stored in the $k$-body Hopfield network $H^{(0)}$ given by Eq.~\eqref{eq:Hfull}. The interaction matrix elements  $J_\chi^{(0)}$ are constructed via $k$-body Hebbian learning of all $M$ patterns as described in Eq.~\eqref{eq:klearning}. The patterns are then local minima of the spectrum [cf. Fig.~\ref{fig1}(a)]. 

\subsubsection{Variational ground-state design}
The structure of the ansatz Hamiltonian~\eqref{eq:Hfull} allows for shifting individual configurations down and up in the energy landscape by Hebbian relearning and unlearning of individual configurations without inducing major shifts in other configurations (cf. Fig.~\ref{fig0}). Such unwanted shifts could be expected in general for all-to-all connected neurons. This observation is used in the following to variationally optimize the energy bandwidth $\Delta_p$ with respect to the energy gap $\Delta_b$ such that the learned patterns become approximately degenerate ground states of the system. 

The initial interaction matrix elements are modified by either applying relearning of patterns or unlearning of bulk configurations such that $\Delta=\Delta_p/\Delta_b$ is minimized. This can be performed by either decreasing the bandwidth $\Delta_p$ or increasing the gap $\Delta_b$. 
In order to decrease the bandwidth $\Delta_p$ pattern $m_\mathrm{max}$ with the highest energy is relearned with small prefactors $\phi_k\ll1$ according to
\begin{equation}
J_\chi^{(1)} = J_\chi^{(0)} -\phi_k\prod_{i=1}^kx_{\chi_i}^{m_\mathrm{max}^{(0)}}.
\end{equation}
Increasing the gap $\Delta_b$ is achieved via Hebbian unlearning of the $r$ lowest-lying bulk configurations $u_b$ with $(b=1,\dots,r$) and small prefactors $\eta_k\ll1$ as
\begin{equation}
J_\chi^{(1')} = J_\chi^{(0)} + \sum_{b=1}^r\eta_k\prod_{i=1}^kx_{\chi_i}^{u_b^{(0)}}.
\end{equation}
In both cases, the re- and unlearning strengths $\phi_k,\eta_k$ are chosen randomly for every re- and unlearning step, respectively. Relearning is applied with probability $p_\mathrm{relearn}$ whereas unlearning is applied with probability $p_\mathrm{unlearn}=1-p_\mathrm{relearn}$.

After every update step, we check the value of $\Delta^{(t)}  = \Delta_p^{(t)}/\Delta_b^{(t)}$, where $t$ counts the number of update steps. If $\Delta^{(t)}>\Delta^*$, where $\Delta^* \ll 1$ is a chosen termination parameter representing a desired $\Delta$, we accept the update with probability $p_A = \min(1,\exp(-\Delta F/ T))$ where the ``temperature''  $T$ is a free optimization parameter and $\Delta F = \Delta_p^{(1)} - \Delta_p^{(0)} - (\Delta_b^{(1)} - \Delta_b^{(0)})$. The update is accepted with certainty if $\Delta F\leq 0$, which corresponds to an improvement of $\Delta^{(t)}$ towards $\Delta^*$. Otherwise the update is rejected. This is iterated until the desired $\Delta^{(t_f)}  = \Delta_p^{(t_f)}/\Delta_b^{(t_f)}\leq\Delta^*$ of Hamiltonian $H^{(t_f)}$  at final update step $t_f>1$ is reached.\\

In principle, the protocol described above requires the energy of all $2^N$ configurations at every update step in order to be able to decide upon our acceptance criterion. A feature of Hopfield networks trained with the Hebbian learning rule is that a configuration that is close to a learned pattern with respect to the Hamming distance is also close in energy (cf. Fig.~\ref{fig1}).  We use this observation in order to make our protocol computationally more efficient. Thus, it is sufficient for our method to utilize only a relatively small subset of configurations, which differ from any stored pattern by a small number of spin flips. 

The fact that patterns close in energy are close in Hamming distance is known as the ``basin of attraction'' property of Hopfield networks. We note that this is another property of Hopfield networks which we use here in a new context. This computational advantage becomes more drastic for bigger system sizes. Since we aim at storing a polynomial number of patterns, the space of relevant configurations grows as $O(N^h)$, whereas the configuration space grows exponentially in $N$. The method is heuristic, which means convergence depends on the requirements on the spectrum and the details of the optimization parameters (e.g., learning rate, ...).

Figure~\ref{fig1} shows typical spectra before and after the iterative ground-state design protocol. Before our iteration, the patterns are local energy minima [cf. Fig.~\ref{fig1}(a) and Fig.~\ref{fig1}(c)], whereas after the iterative process, they are approximate ground states of the spectrum [cf. Fig.~\ref{fig1}(b) and Fig.~\ref{fig1}(d)]. 

\begin{figure}[h]
\begin{centering}
\includegraphics[width=\columnwidth]{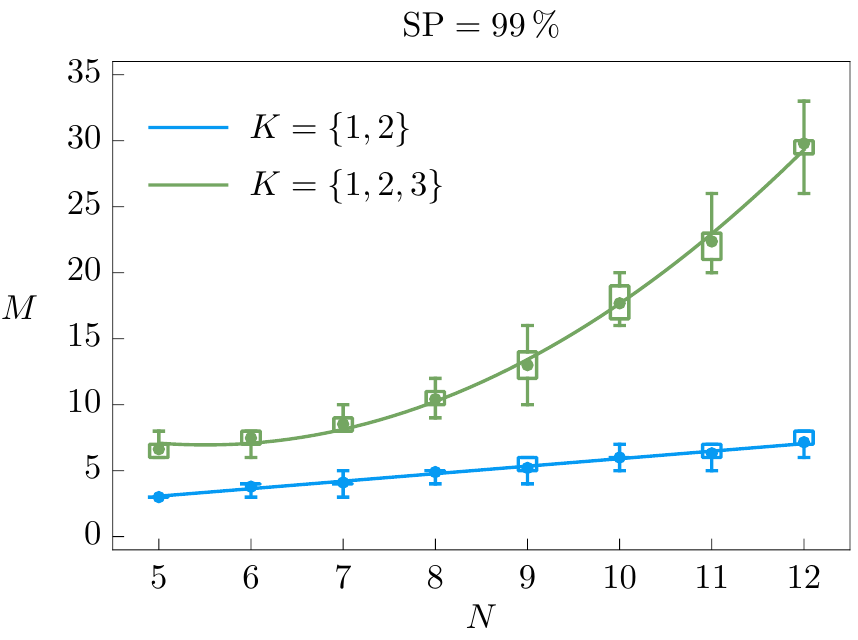}
\par\end{centering}

\protect\caption{Capacity ($\mathrm{SP} = 99\%$) of our protocol for given parameters $\Delta^* = 0.1$, $h=4$, $r = 1$, $p_\mathrm{relearn}=2/3$, and $T = 1$. For every point we have randomly chosen 1000 realizations of $M$ distinct bit strings of length $N$ and tried to design  the desired Hamiltonian for $K =\{1,2\}$ (lower curve) and $K =\{1,2,3\}$ (upper curve). The points in the plot represent the mean of the maximum $M$ (for a given $N$) reached with $\mathrm{SP} = 99\%$ for a significant number of subgroups of size 100 out of 1000 samples. The fit corresponding to the blue (lower) and green (upper) data points is given by $M(N) = 0.57N+0.23$ and $M(N) = 0.53N^2-5.84N+23.02$, respectively.}
\label{fig2}
\end{figure}
\subsection{Capacity of the ground-state design method}
\label{sec:capacity}
An important aspect of neuronal networks is their storage capability~\cite{Amit1985PRL,Amit1985PRA}. Hence, also in our case, it is of interest how many randomly chosen distinct patterns can be stored as ground states utilizing our variational method. To this end, we define the capacity $C$ of our protocol as the maximum number of arbitrary patterns $M$ that can be stored as approximate ground states in a system of size $N$ with a certain success probability (SP) and with respect to a given termination value $\Delta^*$. Thus, $\mathrm{SP}=100\%$ means that the ground-state design method is successful for any combination of $M$ distinct patterns in a system with $N$ spins for a particular $\Delta^*$. Figure~\ref{fig2} gives an estimate of the capacity $C$ of our protocol for $\mathrm{SP} = 99\%$ and $\Delta^* =0.1$. For $K = \{1,2\}$, we find that the capacity increases linearly with the system size $C_{\{1,2\}}= O(N)$  [cf. blue (lower) curve in Fig.~\ref{fig2}]. Including also three-body terms improves the capacity by a factor of $N$ leading to $C_{\{1,2,3\}}= O(N^2)$  [cf. green (upper) curve in Fig.~\ref{fig2}]. The scalings are in good agreement with the analytical upper bound of the storage capacity of Hopfield networks discussed above [cf. Eq.~\eqref{eq:capacity}]. As we will see later, the chosen value $\Delta^*=0.1$ is rather strict compared to typical required values for state preparation.

\section{Programmable Superpositions}
\label{sec:PS}
Now we have the tool at hand to achieve the classical encoding needed for our
goal of creating programed quantum many-body superpositions via quantum
annealing. In the following, we incorporate our method into the
state-preparation protocol of Ref.~\cite{Sieberer2018} and generalize the
latter both to nonperfectly degenerate ground states and higher-order
interactions.

The state preparation protocol of Ref.~\cite{Sieberer2018} can be summarized as follows:
\begin{enumerate}
\item[(i)] \textit{Ground-state design:} Store $M$ classical data bit strings
  $x_n$ in an all-to-all connected spin-glass Hamiltonian $\hat{H}$ (denoted as the
  logical spin model) with the degenerate ground-state manifold spanned by
  $\ket{x_n}$.
\item[(ii)]\textit{Reformulation as a parity-constraint model:} Map $\hat{H}$ to a
  lattice-gauge model according to the LHZ prescription~\cite{Lechner2015}. The
  resulting Hamiltonian $\tilde{H}$ comprises only \emph{local} terms, i.e.,
  local fields and local three- and four-body constraints. This Hamiltonian acts
  on physical qubits which encode the original logical qubits. Thus, logical
  bit strings $x_n$ are translated into physical bit strings $z_n$, representing
  spin configurations of the lattice-gauge model.
\item[(iii)] \textit{State preparation by sweep of a transverse field:} Prepare
  the desired superposition via sweeping a transverse field. This sweep induces
  controlled diabatic transitions within the ground-state manifold. The control
  parameters are given by the constraint strengths of the parity-constraint
  model.
\end{enumerate}

Reference~\cite{Sieberer2018} describes how steps (ii) and (iii) can be
achieved, assuming that step (i) has been accomplished with a two-local
spin-glass Hamiltonian $\hat{H}$, and, in particular, that the states $\ket{x_n}$
are perfectly degenerate ground states of $\hat{H}$. Motivated by the ground-state design method developed in this paper, we now generalize the original
protocol of Ref.~\cite{Sieberer2018} to finite bandwidths of the low-energy
manifold and logical Hamiltonians $\hat{H}$ that include up to three-local
terms. The latter generalization does not affect the dynamical state-preparation
protocol. However, it leads to an increase of the dimensionality of the LHZ
architecture~\cite{Lechner2015} as we will see in the following.

\subsection{Parity-constraint model}
\label{sec:translation}
\begin{figure}[t]
\begin{centering}
\includegraphics[width=\columnwidth]{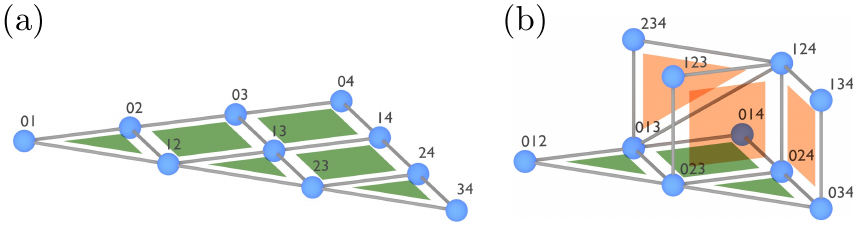}
\end{centering}
\protect\caption{Illustration of (a) 2D and (b) 3D LHZ architectures. Constraints, consisting either of three- or four-body terms, are visualized by shaded triangles or squares. Qubit labels denote the indices of the interaction matrix elements $J_{ij}$ and $J_{ijk}$, respectively, of the two- and three-local spin-glass Hamiltonians.}
\label{fig3}
\end{figure}

The LHZ architecture~\cite{Lechner2015} provides a way of encoding an all-to-all
connected spin-glass Hamiltonian with up to three-local interactions into an
experimentally feasible lattice-gauge representation consisting of only local
fields $\tilde{J}_i$ and local constraints $C_p$. The corresponding Hamiltonian
is of the form
\begin{equation}
\tilde{H} = \tilde{H}_J+\tilde{H}_C,
\end{equation}
with
\begin{equation}
\tilde{H}_J = -\sum_{i=1}^{N_p} \tilde{J}_i\tilde{\sigma}_z^{(i)}, \qquad \tilde{H}_C = \sum_{p}C_p\tilde{S}_p.
\end{equation}
The physical qubits are denoted $\tilde{\sigma}_z^{(i)}$ and $\tilde{S}_p$
is the stabilizer enforcing the constraint labelled by the index $p$
with weight $C_p$. The stabilizers are usually of the form of three- or
four-body $\tilde{\sigma}_z$ terms as described in Ref.~\cite{Lechner2015,Rocchetto2016} and depicted in Fig.~\ref{fig3}. $N_p$ denotes the number of
physical qubits in the LHZ architecture and reflects the number of nonzero
interaction matrix elements present in the logical Hamiltonian. The number of
physical qubits is a function of the $k$-locality, with $N_p = \binom{N}{k}$. At
least $N_p-N+1$ constraints are needed in order to restrict the enlarged Hilbert
space consisting of $2^{N_p}$ states to a low-energy subspace corresponding to the
energies of the $2^N$ configurations of the logical system. Since increasing the
$k$-locality of the spin glass increases the dimensionality of the LHZ
architecture, we focus in the following on the experimentally
realistic~\cite{Glaetzle2017,Leib2016,Puri2017,Chancellor2017} scenarios of two- and three-local logical Hamiltonians.

\subsection{Adiabatic-diabatic state preparation}
The state-preparation protocol developed in Ref.~\cite{Sieberer2018} relies on
an adiabatic-diabatic dynamics within the LHZ encoding. The protocol can be understood as a coherent quantum annealing scheme~\cite{Albash2018} with $M$ degenerate final ground states~\cite{Katzgraber2017,Koenz2018}. In this protocol, one transfers the system prepared in a trivial initial state to the low-energy manifold of the problem Hamiltonian. The time-dependent Hamiltonian describing this protocol is of the form
\begin{equation}
\label{eq:LHZsweep}
\tilde{H}(t) = \delta(t) \tilde{H}_0 + \epsilon(t) \tilde{V},
\end{equation}
with $\tilde{H}_0 = \tilde{H}_J + \tilde{H}_C$ denoting the parity model
encoding of the logical Hamiltonian and $\tilde{V}$ denoting the transverse field
Hamiltonian,
\begin{equation}
\tilde{V} = \sum_{i=1}^{N_p}\tilde{\sigma}^{(i)}_x,
\end{equation}
and the switching functions are given by
\begin{equation}
\label{eq:delta-epsilon}
\delta(t) = t/T, \qquad \epsilon(t) = 1 - t/T.
\end{equation}
Thus, initially $\tilde{H}(0) = \tilde{V}$, and in the
course of the sweep $\tilde{H}(t)$ is transformed into the final Hamiltonian
$\tilde{H}(T) = \tilde{H}_0$.

The evolution of the system's quantum state during a slow sweep from the initial
to the final Hamiltonian depends crucially on the properties of the ground-state
manifold of the final Hamiltonian: For a single nondegenerate ground state of
the final Hamiltonian, the system will follow the instantaneous ground state
adiabatically; Instead, if the ground-state manifold consists of $M$ degenerate
states, the sweep will induce diabatic transitions within the low-energy
manifold spanned by the $M$ lowest-lying instantaneous eigenstates
$ \{\ket{\phi_n(t)}\,|\,n=1,\dots,M\}$ of $\tilde{H}(t)$. Nevertheless, for
long sweep times $T$, transitions out of this manifold are suppressed.

The protocol introduced in Ref.~\cite{Sieberer2018} utilizes the additional
parameters given by the constraint strengths $C_p$ which are introduced in the
LHZ encoding to control the diabatic dynamics within the low-energy manifold,
and thus circumvents the problem of unfair sampling in quantum
annealing~\cite{Katzgraber2017, Koenz2018}. More precisely, the constraints can
be adjusted such that at the end of the sweep the system is in the final-state
$\ket{\psi(T)} = \sum_{n=1}^{M}a_n \ket{z_n}$ with desired probabilities
$\abs{a_n}^2 = p_n^*$.

The separation between diabatic dynamics within the low-energy manifold and
adiabaticity with respect to transitions to the manifold of excited states is
facilitated given exact ground-state degeneracy at the end of the sweep. However, also when the low-energy manifold forms a band of finite width $\Delta_p$, the separation between diabatic and adiabatic dynamics persists as
long as $\Delta_p$ is much larger than the gap $\Delta_b$ separating the
low-energy manifold from the bulk. Below, we discuss the requirements on
$\Delta_p$ and $\Delta_b$ for a specific example.

As transitions out of the low-energy manifold are suppressed for
large enough sweep times, the dynamics is well described by an effective $M$-dimensional theory
by perturbatively decoupling the low-energy subspace from the high-energy
subspace via a Schrieffer-Wolff (SW) transformation (details of the calculation are given in the Appendix). Within the effective
theory, the optimal control parameters can be found by iteratively minimizing
the cost function
\begin{equation}
\label{eq:costfunction}
\Omega(\{a_n\}) = \sum_{n=1}^M \left( |a_n|^2-p_n^* \right)^2,
\end{equation}
where $|a_n|^2=\abs{\langle\phi_n(T)|\psi(T)\rangle}^2$. 
For small system sizes, the full quantum dynamics of the sweep can be calculated
exactly and there is no need to resort to an effective theory.

For degenerate ground states at the end of the sweep, the derivation of
the effective Hamiltonian was discussed in Ref.~\cite{Sieberer2018}. The
generalization to a low-energy manifold with a finite bandwidth $\Delta_p$ is
described in the Appendix.

\subsection{Examples}
\label{subsec:examples}

\begin{figure*}[t]
\begin{centering}
\includegraphics[width=\textwidth]{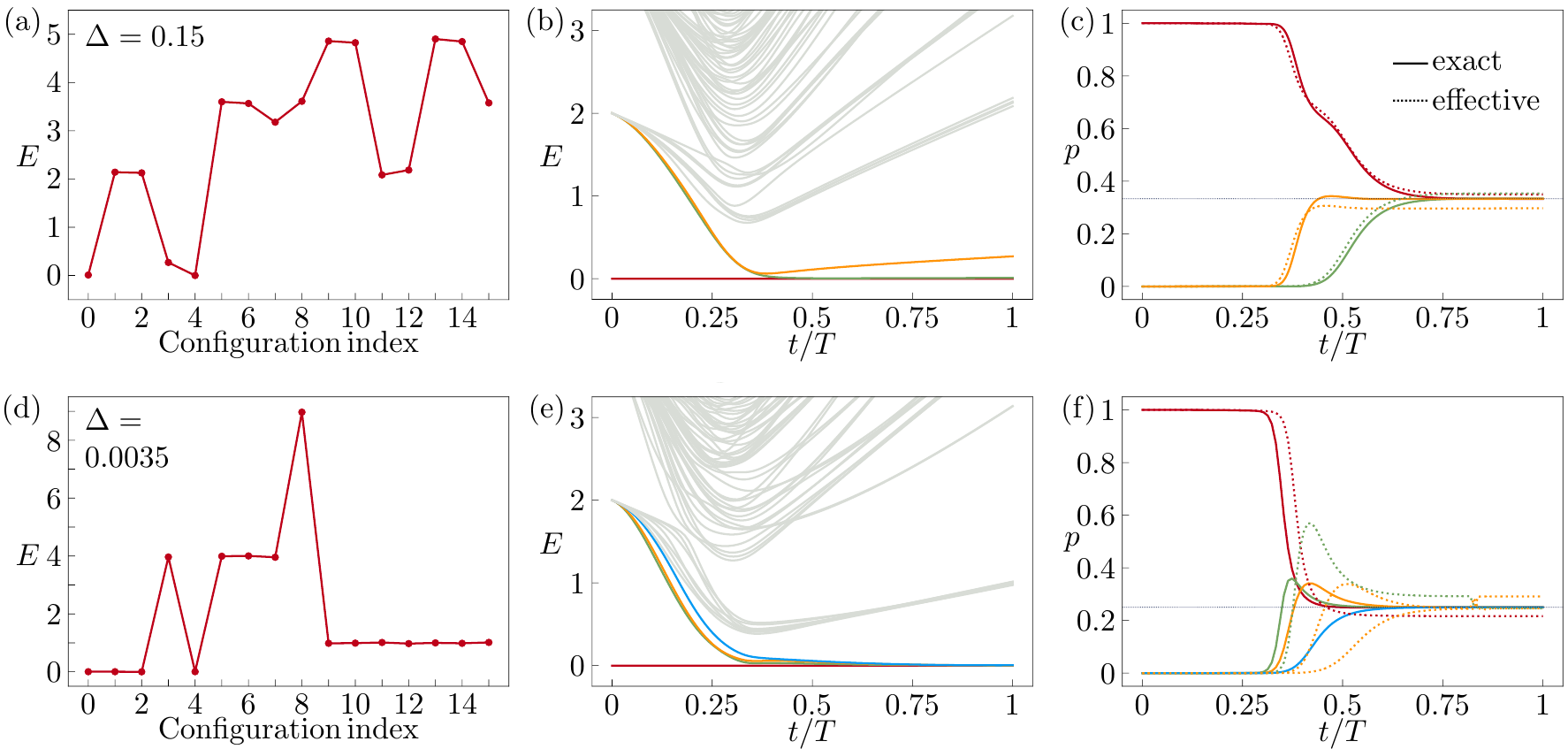}
\par\end{centering}
\protect\caption{Examples of full state-preparation protocol. The upper (lower) row corresponds to the  case of $K=\{1,2\}$ ($K=\{2,3\}$) logical Hamiltonians. (a) Logical spectrum obtained via the iterative ground-state design protocol with $N=4$ and $M=3$ for $\Delta =0.15$ (nondegenerate). (b) Instantaneous energies of the physical system during the sweep described by Eq.~\eqref{eq:LHZsweep}. (c) Overlap $p_n(t) = \abs{\langle\phi_n(t)|\psi(t)\rangle}^2$ of $M$ lowest-energy instantaneous eigenstates with the state of the system during full time evolution for the optimized constraint strengths to obtain uniformly distributed amplitudes $p_n(T)=|a_n|^2=1/M$ (dashed lines: constraint optimization within effective model; solid lines: exact optimization). (d) Logical spectrum obtained via the iterative ground-state design protocol with $N=4$ and $M=4$ for $\Delta =0.0035$ (almost degenerate). (e) Instantaneous energies of the physical system as in (b). (f) Successfully programed amplitudes of $|a_n|^2=1/M$ as in (c).}
\label{fig4}
\end{figure*} 

Having described the full state preparation protocol in detail, we illustrate
the method by two examples with $K=\{1,2\}$ (single- and two-body spin glasses) and $K=\{2,3\}$ (two- and three-body spin glasses).

Figure~\ref{fig3}(a) shows the LHZ architecture for a two-body interacting spin-glass Hamiltonian
$\hat{H}=\sum_{i<j}J_{ij}\hat{\sigma}_z^{(i)}\hat{\sigma}_z^{(j)}$ with $N=5$
logical spins and $N_p=10$ physical qubits. In this two-dimensional parity
architecture, two-body interaction terms in the logical system are represented by a
single physical qubit as
$J_{ij}\hat{\sigma}_z^{(i)}\hat{\sigma}_z^{(j)} \to
J_{ij}\tilde{\sigma}^{(ij)}_z$.
Single-qubit terms can be easily realized by fixing a spin in the logical
system, such that
$J_i\hat{\sigma}_z^{(i)} = J_{0i}\hat{\sigma}_z^{(0)}\hat{\sigma}_z^{(i)}$ with
$\hat{\sigma}_z^{(0)}=1$. The necessary constraint terms of our example can be
realized as local three- and four-body plaquettes [cf. shaded triangles and squares
in Fig.~\ref{fig3}(a)]
\begin{eqnarray}
\label{eq:C2D}
\tilde{H}^{\mathrm{2D}}_C &=&-C_1\tilde{\sigma}^{(01)}_z\tilde{\sigma}^{(02)}_z\tilde{\sigma}^{(12)}_z -C_2\tilde{\sigma}^{(12)}_z\tilde{\sigma}^{(13)}_z\tilde{\sigma}^{(23)}_z\nonumber\\
&&-C_3\tilde{\sigma}^{(23)}_z\tilde{\sigma}^{(24)}_z\tilde{\sigma}^{(34)}_z-C_4\tilde{\sigma}^{(02)}_z\tilde{\sigma}^{(03)}_z\tilde{\sigma}^{(12)}_z\tilde{\sigma}^{(13)}_z\nonumber\\
&&-C_5\tilde{\sigma}^{(13)}_z\tilde{\sigma}^{(14)}_z\tilde{\sigma}^{(23)}_z\tilde{\sigma}^{(24)}_z-C_6\tilde{\sigma}^{(03)}_z\tilde{\sigma}^{(04)}_z\tilde{\sigma}^{(13)}_z\tilde{\sigma}^{(14)}_z,\nonumber\\
\end{eqnarray}
with constraint strengths $C_p>0$.  

A logical three-body interaction
$\hat{H}=\sum_{i<j<k}J_{ijk}\hat{\sigma}_z^{(i)}\hat{\sigma}_z^{(j)}\hat{\sigma}_z^{(k)}$
can be realized in a LHZ architecture with three spatial
dimensions [cf. Fig.~\ref{fig3}(b)]. Again, $N=5$ logical spins correspond to
$N_p=10$ physical qubits. In this case, three-body interaction terms in
the logical system are translated to a single physical qubit as
$J_{ijk}\hat{\sigma}_z^{(i)}\hat{\sigma}_z^{(j)}\hat{\sigma}_z^{(k)}\to
J_{ijk}\tilde{\sigma}^{(ijk)}_z$.
Similar to the 2D case, one can realize single-qubit or two-body interaction
terms by fixing one or two, respectively, logical spins. The
constraint terms corresponding to our example of a two- and three-body
interacting logical system ($\hat{\sigma}_z^{(0)}=1$) are
local three- and four-body plaquettes [cf. shaded triangles and squares in
Fig.~\ref{fig3}(b)]
\begin{eqnarray}
\label{eq:C3D}
\tilde{H}^{\mathrm{3D}}_C &=&-C_1\tilde{\sigma}^{(012)}_z\tilde{\sigma}^{(013)}_z\tilde{\sigma}^{(023)}_z-C_2\tilde{\sigma}^{(023)}_z\tilde{\sigma}^{(024)}_z\tilde{\sigma}^{(034)}_z\nonumber\\
&&-C_3\tilde{\sigma}^{(013)}_z\tilde{\sigma}^{(124)}_z\tilde{\sigma}^{(234)}_z-C_4\tilde{\sigma}^{(013)}_z\tilde{\sigma}^{(014)}_z\tilde{\sigma}^{(023)}_z\tilde{\sigma}^{(024)}_z\nonumber\\
&&-C_5\tilde{\sigma}^{(023)}_z\tilde{\sigma}^{(024)}_z\tilde{\sigma}^{(123)}_z\tilde{\sigma}^{(124)}_z\nonumber\\
&&-C_6\tilde{\sigma}^{(024)}_z\tilde{\sigma}^{(034)}_z\tilde{\sigma}^{(124)}_z\tilde{\sigma}^{(134)}_z,
\end{eqnarray}
where $C_p>0$. 

In reformulating the original spin-glass Hamiltonian in terms of the parity
model, we gain additional tuning knobs provided by the constraint strengths that
allow one to control the quantum dynamics and thus the final amplitudes of the
many-body superpositions.

In the first example, we take a logical system of size $N=4$, $K=\{1,2\}$ and
$M=3$ patterns $x_1=0000$, $x_2 = 0011$, and $x_3 = 0100$ (which correspond to
the configuration indices 0, 3 and 4). The patterns are stored in the low-energy
manifold of the spectrum of the logical Hamiltonian
$\hat{H}=-\sum_iJ_i\hat{\sigma}_z^{(i)}-\sum_{i<j}J_{ij}\hat{\sigma}_z^{(i)}\hat{\sigma}_z^{(j)}$
where the interaction matrix elements $J_1 = 1.00$, $J_2 = 0.37$, $J_3 =0.20$,
$J_4 = 0.21$, $J_{12}=0.35$ , $J_{13} = 0.23$, $J_{14} = 0.22$, $J_{23}=-0.36$,
$J_{24}=-0.37$, and $J_{34}=1.00$ were obtained via our
ground-state design method with $\Delta = 0.15$ [cf. Fig.\ref{fig4}(a)]. This
translates into a two-dimensional LHZ architecture with $N_p=10$ physical qubits
and six constraints as depicted in Fig.~\ref{fig3}(a). The LHZ representation of
the stored patterns is then $\ket{z_1}=\ket{0000000000}$, 
$\ket{z_2}=\ket{0011011110}$, and $\ket{z_3}=\ket{0100100110}$. As described in
Sec.~\ref{sec:translation}, constraints with strengths $C_{1-3}$ are three-body
interactions, whereas the constraints with strengths $C_{4-6}$ are four-body
interactions [cf. Eq.~\eqref{eq:C2D}]. 

In the next step, we use the constraint strengths as control parameters to
prepare a superposition of states $\ket{z_n}$ with target probabilities
$p_n^{*} = 1/M$. In order to find the required constraint strengths $C_p$, we
optimize the cost function Eq.~\eqref{eq:costfunction}. The maximum Hamming
distance between the stored patters is six. Hence, we derive the effective model
in sixth order of perturbation theory, and we find $C_1^\mathrm{eff} = 4.76$,
$C_2^\mathrm{eff}=5.20$, $C_3^\mathrm{eff} = 4.93$, $C_4^\mathrm{eff} = 3.43$,
$C_5^\mathrm{eff} = 2.72$, $C_6^\mathrm{eff}= 2.71$. This is in good agreement
with the exact result $C_1=5.05$, $C_2 = 4.48$, $C_3=5.57$, $C_4 = 3.25$,
$C_5 = 2.12$, $C_6 = 2.74$. Figure~\ref{fig4}(b) shows the time-dependent
spectrum of the Hamiltonian~\eqref{eq:LHZsweep}. Clearly, the low-energy part of
the spectrum of the logical system is correctly reproduced by the LHZ mapping
after optimizing the constraint strengths. Figure~\ref{fig4}(c) shows the
probabilities of the three lowest-lying instantaneous eigenstates of the
time-dependent Hamiltonian~\eqref{eq:LHZsweep}
$p_n(t) = \abs{\langle\phi_n(t)|\psi(t)\rangle}^2$. The exact constraint
strengths generates the desired superposition with
$p_n^*=p_n(T)= 1/3$ [cf. solid lines in Fig.~\ref{fig4}(c)]. The results using $C_p^\mathrm{eff}$ are in good agreement with the results using the exact constraint strengths $C_p$ [cf. dashed lines in Fig.~\ref{fig4}(c)].\\

In the second example, we take a logical system of size $N=4$, $K=\{2,3\}$ and
$M=4$ patterns $x_1=0000$, $x_2 = 0001$, $x_3 = 0010$, and $x_4 = 0100$ (which
correspond to the configuration indices 0, 1, 2 and 4). The patterns are stored
in the low-energy manifold of the spectrum of the logical Hamiltonian
$\hat{H}=-\sum_{i<j}J_{ij}\hat{\sigma}_z^{(i)}\hat{\sigma}_z^{(j)}-\sum_{i<j<k}J_{ijk}\hat{\sigma}_z^{(i)}\hat{\sigma}_z^{(j)}\hat{\sigma}_z^{(k)}$
where the interaction matrix elements $J_{12}=1.00$ , $J_{13} = 0.99$,
$J_{14} = 0.99$, $J_{23}=-0.50$, $J_{24}=-0.50$, $J_{34}=-0.50$, $J_{123}=0.50$
, $J_{124}=0.50$, $J_{134}=0.51$, and $J_{234}=-1.00$ were obtained via our
ground-state design method with $\Delta = 0.0035$ [cf. Fig.\ref{fig4}(d)]. This
translates into a three-dimensional LHZ architecture with $N_p=10$ physical qubits
and six constraints as depicted in Fig.~\ref{fig3}(b). The LHZ representation of
the patterns is then $\ket{z_1}=\ket{0000000000}$, $\ket{z_2}=\ket{0010110111}$,
$\ket{z_3}=\ket{0101011011}$, and $\ket{z_4}=\ket{1001101101}$. Similar to the
two-dimensional case, there are three three-body constraints with strengths
$C_{1-3}$ and three four-body constraints with strengths $C_{4-6}$ [cf.
Eq.~\eqref{eq:C3D}]. However, the constraints are now embedded in a cubic lattice
geometry. 

Again, we use in the second step the constraint strengths as tuning knobs in
order to prepare a superposition of states $\ket{z_n}$ with target
probabilities $p_n^* = 1/M$. Using the effective model up to sixth order of
perturbation theory, we find $C_1^\mathrm{eff} = 5.61$, $C_2^\mathrm{eff}=6.44$,
$C_3^\mathrm{eff} = 3.37$, $C_4^\mathrm{eff} = 6.70$, $C_5^\mathrm{eff} = 2.66$,
$C_6^\mathrm{eff}= 2.60$. Exact optimization yields $C_1=9.01$, $C_2 = 3.31$,
$C_3=4.42$, $C_4 = 9.12$, $C_5 = 2.07$, $C_6 = 2.63$. Figure~\ref{fig4}(e) shows
the time-dependent spectrum of the Hamiltonian~\eqref{eq:LHZsweep}. Clearly, the
low-energy spectrum of the logical system is correctly reproduced by the LHZ
mapping after optimizing the constraint strengths. Figure~\ref{fig4}(f) shows
the probabilities $p_n(t)$ of the three lowest-lying instantaneous eigenstates
of the time-dependent Hamiltonian~\eqref{eq:LHZsweep}. Using the exact
constraint strengths generates the desired superposition with $p_n^*= 1/4$ [solid
lines in Fig.~\ref{fig4}(f)]. The results using $C_p^\mathrm{eff}$ are in good
agreement with the results using the exact constraint strengths $C_p$
[cf. dashed lines in Fig.~\ref{fig4}(f)].

\begin{figure*}[t]
\begin{centering}
\includegraphics[width=\textwidth]{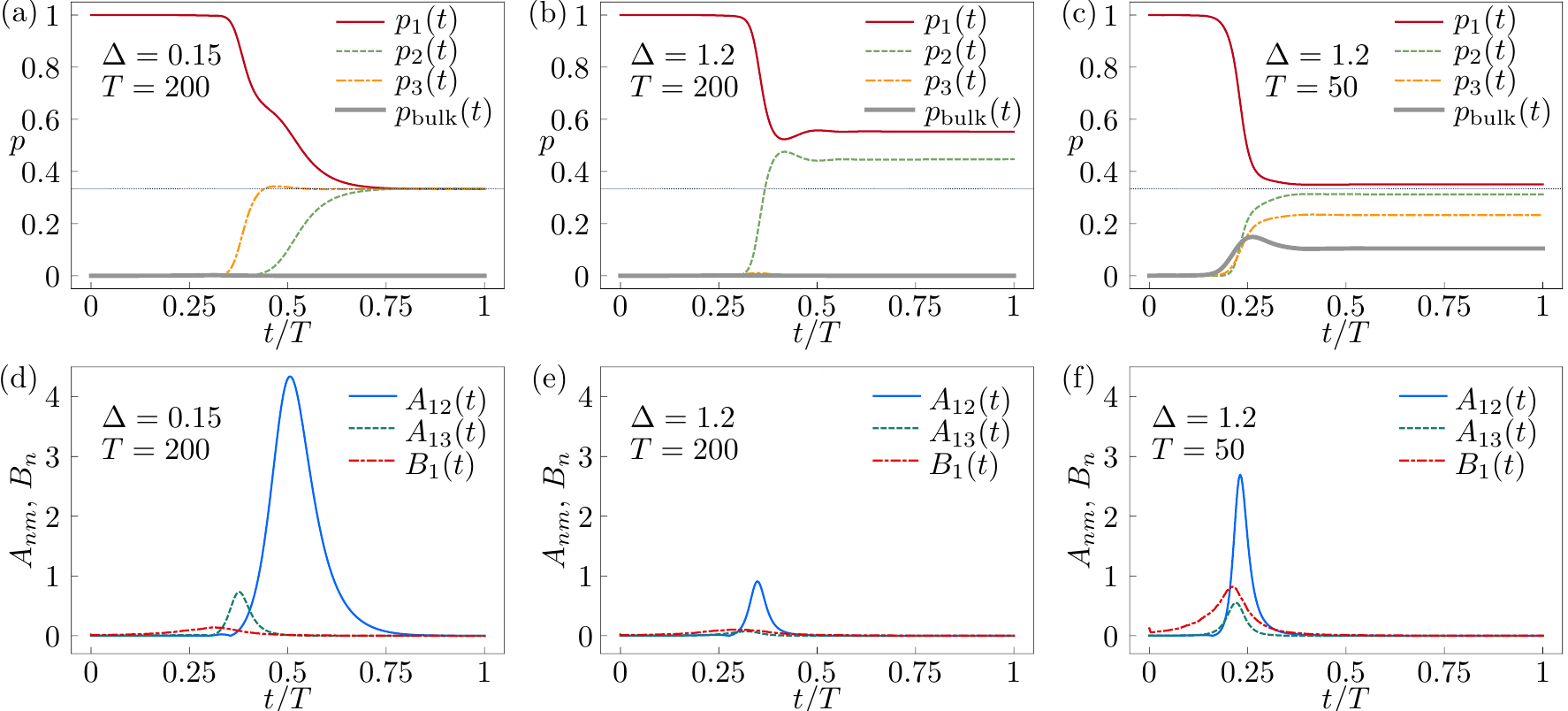}
\par\end{centering}
\protect\caption{Adiabatic and diabatic dynamics for different combinations of
  $\Delta = \Delta_p/\Delta_b$ and $T$ . The upper panels [(a)-(c)] show the overlaps
  $p_n(t) = \abs{\langle\phi_n(t)|\psi(t)\rangle}^2$ of the $M=3$ instantaneous
  eigenstates which form the low-energy manifold with the state of the system
  $\ket{\psi(t)}$ during the sweep of the transverse field and the total
  population of the bulk
  $p_\mathrm{bulk} = \sum_{n>3}\abs{\langle\phi_n(t)|\psi(t)\rangle}^2$. The
  lower panels [(d)-(f)] show the parameters $A_{12}(t)$, $A_{13}(t)$, and $B_1(t)$ as
  defined in the main text. The optimization is successful if $p_n(T)=1/M$ and
  $p_\mathrm{bulk}(T)=0$ at the end of the sweep. (a) and (d) For the example of
  Fig.~\ref{fig4}(a-c), the parameters $A_{12}(t)$,$A_{13}(t)$ defined in
  Eq.~\eqref{eq:A-nm} are strongly peaked with maximum values $\gtrsim 1$ during
  short periods of population transfer from levels $1 \to 2$ and $1 \to 3$. In
  contrast, $B_{1}(t) \ll 1$ at all times. (b) and (d) Optimization of the same example
  as in (a) does not converge for $\Delta=1.2$ since necessary diabatic transitions
  within the low-energy manifold (here $1 \to 3$) are absent $(A_{13}(t)\ll1)$.
  (e) and (f) Trying to compensate the situation of (b) by decreasing $T$ leads to
  diabatic transitions into the bulk $B_{1}(t)\gtrsim 1$. This results in
  $p_1(T)+p_2(T)+p_3(T)<1$ and $p_{\mathrm{bulk}}(T) > 0$ at the end of the
  sweep.}
\label{fig5}
\end{figure*}
\subsubsection*{Interplay between adiabatic and diabatic dynamics}

In the following, we discuss the relevance of diabatic and adiabatic dynamics in our state-preparation protocol for the choice of a suitable  value $\Delta$ for successful optimization. The maximum permissible value of
$\Delta_p$ for successful state preparation is determined by the condition that
it should be possible to induce diabatic transitions within the instantaneous
low-energy manifold; on the other hand, the dynamics must be adiabatic with
respect to transitions out of the instantaneous low-energy manifold. The latter
requirement poses a constraint on the minimum allowed value of the gap
$\Delta_b$ between the low-energy manifold and the lowest-lying bulk state.

A measure for the adiabaticity of the dynamics with respect to transitions
between instantaneous eigenstates $\ket{\phi_n(t)}$ and $\ket{\phi_m(t)}$ with
corresponding energies $E_n(t)$ and $E_m(t)$, respectively, is given
by~\cite{Albash2018,DeGrandi2010,Amin2009}
\begin{equation}
  \label{eq:A-nm}
  A_{nm}(t) = \abs{\frac{\braket{\phi_n(t) | \dot{\tilde{H}}(t) |
        \phi_m(t)}}{[E_n(t) - E_m(t)]^2}}.
\end{equation}
To enable the transfer of populations between states $\ket{\phi_n(t)}$ and
$\ket{\phi_m(t)}$ within the low-energy manifold, we require $A_{nm} \gtrsim 1$,
while $A_{nm} \ll 1$ should be maintained at all times if one of the states
belongs to the bulk of excited states. 

For the example of Fig.~\ref{fig4}(a) the parameters
$A_{12}(t)$ and $A_{13}(t)$ are shown in Fig.~\ref{fig5}(d). The peaks in these parameters are in direct correspondence
with the rather short periods of transfers of population from levels $1 \to 2$
and $1 \to 3$, as can be seen in Fig.~\ref{fig5}(a). To quantify
leakage of population out of the low-energy manifold into the bulk, we consider
the quantities $B_n(t)=\sum_{m>n}A_{nm}(t)$. As also shown in Fig.~\ref{fig5}(d), $B_1(t) \ll 1$ during the entire sweep, which indicates adiabaticity of the dynamics with respect to transitions to the bulk.

The key tuning parameters to achieve the required conditions of adibaticity and
diabaticity are the width of the low-energy manifold $\Delta_p$, the bulk gap
$\Delta_b$, and the total sweep time $T$. Intuitively, a small value of
$\Delta_p$ enables diabatic transitions within the low-energy
manifold. Increasing the value of $\Delta_p$ has to be compensated by decreasing
$T$. However, this also makes unwanted transitions to the bulk more likely and
thus requires an even larger value of $\Delta_b$. Finding a suitable parameter
regime to carry out the state preparation thus requires
$\Delta = \Delta_p/\Delta_b \ll 1$. For the present example, we found
$\Delta = 0.15$ to be sufficient. Making a general prediction for the required
value of $\Delta$ is difficult due the interplay with other problem-specific
parameters, such as the Hamming distances between the final states within the
low-energy manifold. Indeed, for a given Hopfield Hamiltonian with parameters
$\Delta_p$ and $\Delta_b$, the constraint strengths $C_p$ and sweep time $T$
have to be found from an optimization as described above.

We can confirm the validity of the above intuition by attempting the same
optimization task as in the example of Fig.~\ref{fig4}(a), but for a critical
value of $\Delta\approx 1$, which corresponds to an increase in $\Delta_p$ and a
decrease in $\Delta_b$. If we keep the same value of the run time $T$, necessary
diabatic transitions within the low energy manifold are absent
[cf.~\ref{fig5}(e)]. As a result, the optimization of the constraint strengths
does not converge, and the sweep of the transverse field in the
Hamiltonian~\eqref{eq:LHZsweep} fails to prepare the desired superposition state
[cf. Fig.~\ref{fig5}(b)]. One can try to compensate the increase in $\Delta_p$
by decreasing $T$. However, this also leads to a nonconverging optimization
[cf. Fig.~\ref{fig5}(c)]. As shown in Fig.~\ref{fig5}(f), the insufficient
energetic separation between the low-energy subspace and the bulk enables
diabatic transitions into the bulk.

\section{Discussion}
\label{sec:conclusion}

In the present paper, we propose a variational method based on $k$-local Hopfield
networks, which allows to design the spectrum of an all-to-all connected Ising
Hamiltonian such that a polynomial number of configurations are approximately
degenerate ground states of the system. An analysis of the capacity of this
approach reveals that it matches the general capacity of (nonperturbed)
$k$-local Hopfield networks.

These findings allow us to complete and extend the state preparation protocol of
Ref.~\cite{Sieberer2018}, which only needs a polynomial number of qubits and can
be implemented in state-of-the-art experiments, e.g., neutral
atoms~\cite{Glaetzle2017} or superconducting
qubits~\cite{Leib2016,Puri2017,Chancellor2017}. In particular, we find that
perfect degeneracy is not necessary for the state-preparation protocol of
Ref.~\cite{Sieberer2018}.

The full state-preparation method described here can be seen as hybrid approach utilizing a classical higher-order Hopfield network
combined with a new quantum recall phase providing superpositions of stored
patterns. 

Possible extensions of our approach include the use of higher-order stabelizers as proposed in Ref.~\cite{Rocchetto2016}. Also, the individual phases may be controlled utilizing phase-dependent cost functions and inhomogeneous driver Hamiltonians, which will be subject of future work. 

We hope that this paper can be useful for applications in quantum machine learning, which benefit from data provided as superpositions
\cite{HHL2009,Rebentrost2014,Lloyd2014,Biamonte2017,Lloyd2013,Dunjko2018}.

\section{Acknowledgements}
We thank M. Leib and K. Ender for valuable discussions. Research was funded by the Austrian Science Fund (FWF) through a START Grant under Project No. Y1067-N27, the Hauser-Raspe Foundation, and by the ERC through the synergy Grant UQUAM.

\appendix
\section{Effective Hamiltonian}
\label{sec:methods}

In this section, we summarize how the effective Hamiltonian
$\tilde{H}_\mathrm{eff}$ (used in Sec.~\ref{subsec:examples}) can be obtained
via SW pertubation theory. We follow the notation of Ref.~\cite{Bravyi2011}. The
case of perfectly degenerate ground states of the logical system was discussed
in Ref.~\cite{Sieberer2018}. Here, we generalize this approach by allowing for a
nondegenerate low-energy manifold and higher-order interactions. General
statements made in in Ref.~\cite{Sieberer2018} regarding the structure of the
perturbative expansion also apply here. In the following, we focus mainly on the
technical differences appearing due to nondegenerate low-energy states and
higher-order interactions.

For long times $t$ with $(T - t)/T \ll 1$, the driver Hamiltonian
$\tilde{V}(t) = \epsilon(t)\tilde{V}$ can be regarded as a perturbation to the
Hamiltonian $\tilde{H}_0(t) = \delta(t)\tilde{H}_0$ since
$\epsilon(t)\ll\delta(t)$. Thus, we can treat $\epsilon(t)$ as the expansion
parameter. Defining the projector on the low-energy manifold
$P = \sum_{n= 1}^M\ket{z_n}\bra{z_n}$ and the projector on the excited-state
space $Q =\mathbbm{1} - P$ of $\tilde{H}_0$, the general structure of the
expansion has the form
\begin{equation}
  H_\mathrm{eff}(t) = \delta(t) \tilde{H}_0P + \epsilon(t) P\tilde{V}P + \sum_{n=2}^\infty
  \frac{\epsilon(t)^n}{\delta(t)^{n - 1}} H_{\mathrm{eff}, n},
\end{equation}
where the operators $H_{\mathrm{eff},n}$ are time independent. Since the goal of
the SW transformation is to bring the Hamiltonian to block-diagonal form, it is
useful to introduce the following superoperators:
\begin{equation}
\mathcal{D}(X) = PXP + QXQ, \qquad \mathcal{O}(X) = PXQ + QXP.
\end{equation}
Every operator can be decomposed into block-diagonal and block-off-diagonal components, which results in
\begin{equation}
\tilde{V} = V_\mathrm{d} + V_\mathrm{od},\qquad   V_\mathrm{d} =\mathcal{D}(\tilde{V}),\qquad V_\mathrm{od} =\mathcal{O}(\tilde{V}).
\end{equation}
We define another superoperator $\mathcal{L}$ by
\begin{equation}
  \mathcal{L}(X) = \sum_{i,j}\frac{\bra{i}\mathcal{O}(X)\ket{j}}{E_i-E_j}\ket{i}\bra{j}-\mathrm{H.c.}.
\end{equation}
States $\ket{i}$ and the corresponding energies $E_i$ denote the (nondegenerate) eigenstates of the unperturbed Hamiltonian $\tilde{H}_0$ representing the low energy manifold, whereas $\ket{j}$ are the eigenstates representing the bulk states with energies $E_j$. 

The expressions for $H_{\mathrm{eff},n}$ for $n\leq 4$ are given by
\begin{equation}
\begin{split}
H_{\mathrm{eff},2} &=  \frac{1}{2}P\left[S_1,V_\mathrm{od}\right]P,\qquad H_{\mathrm{eff},3}  = \frac{1}{2}P[S_2, V_\mathrm{od}]P,\\
H_{\mathrm{eff},4} &=   \frac{1}{2}P[S_3,V_\mathrm{od}]P  -\frac{1}{24}P[S_1,[S_1,[S_1,V_\mathrm{od}]]]P,
\end{split}
\end{equation}
with the operators $S_{i}$ defined as
\begin{eqnarray}
\begin{split}
S_1 &= \mathcal{L}(V_\mathrm{od}),\qquad S_2 =  -\mathcal{L}([V_\mathrm{d},S_1])\\
S_3 & =  -\mathcal{L}([V_\mathrm{d},S_2]).
\end{split}
\end{eqnarray}
Higher-order expressions can be obtained by following the iterative procedure described in Ref~\cite{Bravyi2011}.


\begin{thebibliography}{46}%
\makeatletter
\providecommand \@ifxundefined [1]{%
 \@ifx{#1\undefined}
}%
\providecommand \@ifnum [1]{%
 \ifnum #1\expandafter \@firstoftwo
 \else \expandafter \@secondoftwo
 \fi
}%
\providecommand \@ifx [1]{%
 \ifx #1\expandafter \@firstoftwo
 \else \expandafter \@secondoftwo
 \fi
}%
\providecommand \natexlab [1]{#1}%
\providecommand \enquote  [1]{``#1''}%
\providecommand \bibnamefont  [1]{#1}%
\providecommand \bibfnamefont [1]{#1}%
\providecommand \citenamefont [1]{#1}%
\providecommand \href@noop [0]{\@secondoftwo}%
\providecommand \href [0]{\begingroup \@sanitize@url \@href}%
\providecommand \@href[1]{\@@startlink{#1}\@@href}%
\providecommand \@@href[1]{\endgroup#1\@@endlink}%
\providecommand \@sanitize@url [0]{\catcode `\\12\catcode `\$12\catcode
  `\&12\catcode `\#12\catcode `\^12\catcode `\_12\catcode `\%12\relax}%
\providecommand \@@startlink[1]{}%
\providecommand \@@endlink[0]{}%
\providecommand \url  [0]{\begingroup\@sanitize@url \@url }%
\providecommand \@url [1]{\endgroup\@href {#1}{\urlprefix }}%
\providecommand \urlprefix  [0]{URL }%
\providecommand \Eprint [0]{\href }%
\providecommand \doibase [0]{http://dx.doi.org/}%
\providecommand \selectlanguage [0]{\@gobble}%
\providecommand \bibinfo  [0]{\@secondoftwo}%
\providecommand \bibfield  [0]{\@secondoftwo}%
\providecommand \translation [1]{[#1]}%
\providecommand \BibitemOpen [0]{}%
\providecommand \bibitemStop [0]{}%
\providecommand \bibitemNoStop [0]{.\EOS\space}%
\providecommand \EOS [0]{\spacefactor3000\relax}%
\providecommand \BibitemShut  [1]{\csname bibitem#1\endcsname}%
\let\auto@bib@innerbib\@empty
\bibitem [{\citenamefont {Cirac}\ and\ \citenamefont
  {Zoller}(2012)}]{Cirac2012}%
  \BibitemOpen
  \bibfield  {author} {\bibinfo {author} {\bibfnamefont {J.~I.}\
  \bibnamefont {Cirac}}\ and\ \bibinfo {author} {\bibfnamefont {P.}\
  \bibnamefont {Zoller}},\ }\bibfield  {title} {{\bibinfo {title}
  {Goals and opportunities in quantum simulation},}\ }\href
  {http://dx.doi.org/10.1038/nphys2275} {\bibfield  {journal} {\bibinfo
  {journal} {Nat. Phys.}\ }\textbf {\bibinfo {volume} {8}},\ \bibinfo
  {pages} {264} (\bibinfo {year} {2012})}\BibitemShut {NoStop}%
\bibitem [{\citenamefont {Georgescu}\ \emph {et~al.}(2014)\citenamefont
  {Georgescu}, \citenamefont {Ashhab},\ and\ \citenamefont
  {Nori}}]{Georgescu2014}%
  \BibitemOpen
  \bibfield  {author} {\bibinfo {author} {\bibfnamefont {I.~M.}\ \bibnamefont
  {Georgescu}}, \bibinfo {author} {\bibfnamefont {S.}~\bibnamefont {Ashhab}}, \
  and\ \bibinfo {author} {\bibfnamefont {F.}\ \bibnamefont {Nori}},\
  }\bibfield  {title} {\enquote {\bibinfo {title} {Quantum simulation},}\
  }\href {\doibase 10.1103/RevModPhys.86.153} {\bibfield  {journal} {\bibinfo
  {journal} {Rev. Mod. Phys.}\ }\textbf {\bibinfo {volume} {86}},\ \bibinfo
  {pages} {153} (\bibinfo {year} {2014})}\BibitemShut {NoStop}%
\bibitem [{\citenamefont {DiCarlo}\ \emph {et~al.}(2010)\citenamefont
  {DiCarlo}, \citenamefont {Reed}, \citenamefont {Sun}, \citenamefont
  {Johnson}, \citenamefont {Chow}, \citenamefont {Gambetta}, \citenamefont
  {Frunzio}, \citenamefont {Girvin}, \citenamefont {Devoret},\ and\
  \citenamefont {Schoelkopf}}]{DiCarlo2010}%
  \BibitemOpen
  \bibfield  {author} {\bibinfo {author} {\bibfnamefont {L.}~\bibnamefont
  {DiCarlo}}, \bibinfo {author} {\bibfnamefont {M.~D.}\ \bibnamefont {Reed}},
  \bibinfo {author} {\bibfnamefont {L.}~\bibnamefont {Sun}}, \bibinfo {author}
  {\bibfnamefont {B.~R.}\ \bibnamefont {Johnson}}, \bibinfo {author}
  {\bibfnamefont {J.~M.}\ \bibnamefont {Chow}}, \bibinfo {author}
  {\bibfnamefont {J.~M.}\ \bibnamefont {Gambetta}}, \bibinfo {author}
  {\bibfnamefont {L.}~\bibnamefont {Frunzio}}, \bibinfo {author} {\bibfnamefont
  {S.~M.}\ \bibnamefont {Girvin}}, \bibinfo {author} {\bibfnamefont {M.~H.}\
  \bibnamefont {Devoret}}, \ and\ \bibinfo {author} {\bibfnamefont {R.~J.}\
  \bibnamefont {Schoelkopf}},\ }\bibfield  {title} {\enquote {\bibinfo {title}
  {Preparation and measurement of three-qubit entanglement in a superconducting
  circuit},}\ }\href {http://dx.doi.org/10.1038/nature09416} {\bibfield
  {journal} {\bibinfo  {journal} {Nature (London)}\ }\textbf {\bibinfo {volume} {467}},\
  \bibinfo {pages} {574} (\bibinfo {year} {2010})}\BibitemShut {NoStop}%
\bibitem [{\citenamefont {Bernien}\ \emph {et~al.}(2017)\citenamefont
  {Bernien}, \citenamefont {Schwartz}, \citenamefont {Keesling}, \citenamefont
  {Levine}, \citenamefont {Omran}, \citenamefont {Pichler}, \citenamefont
  {Choi}, \citenamefont {Zibrov}, \citenamefont {Endres}, \citenamefont
  {Greiner}, \citenamefont {Vuleti{\'c}},\ and\ \citenamefont
  {Lukin}}]{Bernien2017}%
  \BibitemOpen
  \bibfield  {author} {\bibinfo {author} {\bibfnamefont {H.}\ \bibnamefont
  {Bernien}}, \bibinfo {author} {\bibfnamefont {S.}\ \bibnamefont
  {Schwartz}}, \bibinfo {author} {\bibfnamefont {A.}\ \bibnamefont
  {Keesling}}, \bibinfo {author} {\bibfnamefont {H.}\ \bibnamefont
  {Levine}}, \bibinfo {author} {\bibfnamefont {A.}\ \bibnamefont {Omran}},
  \bibinfo {author} {\bibfnamefont {H.}\ \bibnamefont {Pichler}}, \bibinfo
  {author} {\bibfnamefont {S.}\ \bibnamefont {Choi}}, \bibinfo {author}
  {\bibfnamefont {A.~S.}\ \bibnamefont {Zibrov}}, \bibinfo {author}
  {\bibfnamefont {M.}\ \bibnamefont {Endres}}, \bibinfo {author}
  {\bibfnamefont {M.}\ \bibnamefont {Greiner}}, \bibinfo {author}
  {\bibfnamefont {V.}\ \bibnamefont {Vuleti{\'c}}}, \ and\ \bibinfo
  {author} {\bibfnamefont {M.~D.}\ \bibnamefont {Lukin}},\ }\bibfield
  {title} {\enquote {\bibinfo {title} {Probing many-body dynamics on a 51-atom
  quantum simulator},}\ }\href {http://dx.doi.org/10.1038/nature24622}
  {\bibfield  {journal} {\bibinfo  {journal} {Nature (London)}\ }\textbf {\bibinfo
  {volume} {551}},\ \bibinfo {pages} {579} (\bibinfo {year}
  {2017})}\BibitemShut {NoStop}%
\bibitem [{\citenamefont {Raimond}\ \emph {et~al.}(2001)\citenamefont
  {Raimond}, \citenamefont {Brune},\ and\ \citenamefont
  {Haroche}}]{Raimond2001}%
  \BibitemOpen
  \bibfield  {author} {\bibinfo {author} {\bibfnamefont {J.~M.}\ \bibnamefont
  {Raimond}}, \bibinfo {author} {\bibfnamefont {M.}~\bibnamefont {Brune}}, \
  and\ \bibinfo {author} {\bibfnamefont {S.}~\bibnamefont {Haroche}},\
  }\bibfield  {title} {\enquote {\bibinfo {title} {Manipulating quantum
  entanglement with atoms and photons in a cavity},}\ }\href {\doibase
  10.1103/RevModPhys.73.565} {\bibfield  {journal} {\bibinfo  {journal} {Rev.
  Mod. Phys.}\ }\textbf {\bibinfo {volume} {73}},\ \bibinfo {pages} {565}
  (\bibinfo {year} {2001})}\BibitemShut {NoStop}%
\bibitem [{\citenamefont {Harrow}\ \emph {et~al.}(2009)\citenamefont {Harrow},
  \citenamefont {Hassidim},\ and\ \citenamefont {Lloyd}}]{HHL2009}%
  \BibitemOpen
  \bibfield  {author} {\bibinfo {author} {\bibfnamefont {A.~W.}\ \bibnamefont
  {Harrow}}, \bibinfo {author} {\bibfnamefont {A.}\ \bibnamefont
  {Hassidim}}, \ and\ \bibinfo {author} {\bibfnamefont {S.}\ \bibnamefont
  {Lloyd}},\ }\bibfield  {title} {\enquote {\bibinfo {title} {Quantum algorithm
  for linear systems of equations},}\ }\href {\doibase
  10.1103/PhysRevLett.103.150502} {\bibfield  {journal} {\bibinfo  {journal}
  {Phys. Rev. Lett.}\ }\textbf {\bibinfo {volume} {103}},\ \bibinfo {pages}
  {150502} (\bibinfo {year} {2009})}\BibitemShut {NoStop}%
\bibitem [{\citenamefont {Rebentrost}\ \emph {et~al.}(2014)\citenamefont
  {Rebentrost}, \citenamefont {Mohseni},\ and\ \citenamefont
  {Lloyd}}]{Rebentrost2014}%
  \BibitemOpen
  \bibfield  {author} {\bibinfo {author} {\bibfnamefont {P.}\ \bibnamefont
  {Rebentrost}}, \bibinfo {author} {\bibfnamefont {M.}\ \bibnamefont
  {Mohseni}}, \ and\ \bibinfo {author} {\bibfnamefont {S.}\ \bibnamefont
  {Lloyd}},\ }\bibfield  {title} {\enquote {\bibinfo {title} {Quantum support
  vector machine for big data classification},}\ }\href {\doibase
  10.1103/PhysRevLett.113.130503} {\bibfield  {journal} {\bibinfo  {journal}
  {Phys. Rev. Lett.}\ }\textbf {\bibinfo {volume} {113}},\ \bibinfo {pages}
  {130503} (\bibinfo {year} {2014})}\BibitemShut {NoStop}%
\bibitem [{\citenamefont {Lloyd}\ \emph {et~al.}(2014)\citenamefont {Lloyd},
  \citenamefont {Mohseni},\ and\ \citenamefont {Rebentrost}}]{Lloyd2014}%
  \BibitemOpen
  \bibfield  {author} {\bibinfo {author} {\bibfnamefont {S.}\ \bibnamefont
  {Lloyd}}, \bibinfo {author} {\bibfnamefont {M.}\ \bibnamefont {Mohseni}},
  \ and\ \bibinfo {author} {\bibfnamefont {P.}\ \bibnamefont
  {Rebentrost}},\ }\bibfield  {title} {\enquote {\bibinfo {title} {Quantum
  principal component analysis},}\ }\href {http://dx.doi.org/10.1038/nphys3029}
  {\bibfield  {journal} {\bibinfo  {journal} {Nat. Phys.}\ }\textbf
  {\bibinfo {volume} {10}},\ \bibinfo {pages} {631} (\bibinfo {year}
  {2014})}\BibitemShut {NoStop}%
\bibitem [{\citenamefont {Biamonte}\ \emph {et~al.}(2017)\citenamefont
  {Biamonte}, \citenamefont {Wittek}, \citenamefont {Pancotti}, \citenamefont
  {Rebentrost}, \citenamefont {Wiebe},\ and\ \citenamefont
  {Lloyd}}]{Biamonte2017}%
  \BibitemOpen
  \bibfield  {author} {\bibinfo {author} {\bibfnamefont {J.}\ \bibnamefont
  {Biamonte}}, \bibinfo {author} {\bibfnamefont {P.}\ \bibnamefont
  {Wittek}}, \bibinfo {author} {\bibfnamefont {N.}\ \bibnamefont
  {Pancotti}}, \bibinfo {author} {\bibfnamefont {P.}\ \bibnamefont
  {Rebentrost}}, \bibinfo {author} {\bibfnamefont {N.}\ \bibnamefont
  {Wiebe}}, \ and\ \bibinfo {author} {\bibfnamefont {S.}\ \bibnamefont
  {Lloyd}},\ }\bibfield  {title} {\enquote {\bibinfo {title} {Quantum machine
  learning},}\ }\href {http://dx.doi.org/10.1038/nature23474} {\bibfield
  {journal} {\bibinfo  {journal} {Nature (London)}\ }\textbf {\bibinfo {volume} {549}},\
  \bibinfo {pages} {195} (\bibinfo {year} {2017})}\BibitemShut {NoStop}%
\bibitem [{\citenamefont {Lloyd}\ \emph {et~al.}()\citenamefont {Lloyd},
  \citenamefont {Mohseni},\ and\ \citenamefont {Rebentrost}}]{Lloyd2013}%
  \BibitemOpen
  \bibfield  {author} {\bibinfo {author} {\bibfnamefont {S.}\ \bibnamefont
  {Lloyd}}, \bibinfo {author} {\bibfnamefont {M.}\ \bibnamefont {Mohseni}},
  \ and\ \bibinfo {author} {\bibfnamefont {P.}\ \bibnamefont
  {Rebentrost}},\ }\bibfield  {title} {\enquote {\bibinfo {title} {Quantum
  algorithms for supervised and unsupervised machine learning},}\ }\href
  {https://arxiv.org/abs/1307.0411} {\bibinfo  {journal} {arXiv:1307.0411}}\BibitemShut {NoStop}%
\bibitem [{\citenamefont {Dunjko}\ and\ \citenamefont
  {Briegel}(2018)}]{Dunjko2018}%
  \BibitemOpen
\bibfield  {journal} {  }\bibfield  {author} {\bibinfo {author} {\bibfnamefont
  {V.}\ \bibnamefont {Dunjko}}\ and\ \bibinfo {author} {\bibfnamefont
  {H.~J.}\ \bibnamefont {Briegel}},\ }\bibfield  {title} {\enquote {\bibinfo
  {title} {Machine learning \& artificial intelligence in the quantum domain: A
  review of recent progress},}\ }\href
  {https://iopscience.iop.org/article/10.1088/1361-6633/aab406} {\bibfield  {journal} {\bibinfo
  {journal} {Rep. Prog. Phys.}\ }\textbf {\bibinfo {volume}
  {81}},\ \bibinfo {pages} {074001} (\bibinfo {year} {2018})}\BibitemShut
  {NoStop}%
\bibitem [{\citenamefont {Giovannetti}\ \emph
  {et~al.}(2008{\natexlab{a}})\citenamefont {Giovannetti}, \citenamefont
  {Lloyd},\ and\ \citenamefont {Maccone}}]{Giovanetti2008a}%
  \BibitemOpen
  \bibfield  {author} {\bibinfo {author} {\bibfnamefont {V.}\
  \bibnamefont {Giovannetti}}, \bibinfo {author} {\bibfnamefont {S.}\
  \bibnamefont {Lloyd}}, \ and\ \bibinfo {author} {\bibfnamefont {L.}\
  \bibnamefont {Maccone}},\ }\bibfield  {title} {\enquote {\bibinfo {title}
  {Quantum random access memory},}\ }\href {\doibase
  10.1103/PhysRevLett.100.160501} {\bibfield  {journal} {\bibinfo  {journal}
  {Phys. Rev. Lett.}\ }\textbf {\bibinfo {volume} {100}},\ \bibinfo {pages}
  {160501} (\bibinfo {year} {2008}{\natexlab{a}})}\BibitemShut {NoStop}%
\bibitem [{\citenamefont {Giovannetti}\ \emph
  {et~al.}(2008{\natexlab{b}})\citenamefont {Giovannetti}, \citenamefont
  {Lloyd},\ and\ \citenamefont {Maccone}}]{Giovanetti2008b}%
  \BibitemOpen
  \bibfield  {author} {\bibinfo {author} {\bibfnamefont {V.}\
  \bibnamefont {Giovannetti}}, \bibinfo {author} {\bibfnamefont {S.}\
  \bibnamefont {Lloyd}}, \ and\ \bibinfo {author} {\bibfnamefont {L.}\
  \bibnamefont {Maccone}},\ }\bibfield  {title} {\enquote {\bibinfo {title}
  {Architectures for a quantum random access memory},}\ }\href {\doibase
  10.1103/PhysRevA.78.052310} {\bibfield  {journal} {\bibinfo  {journal} {Phys.
  Rev. A}\ }\textbf {\bibinfo {volume} {78}},\ \bibinfo {pages} {052310}
  (\bibinfo {year} {2008}{\natexlab{b}})}\BibitemShut {NoStop}%
\bibitem [{\citenamefont {Preskill}(2018)}]{Preskill2018}%
  \BibitemOpen
  \bibfield  {author} {\bibinfo {author} {\bibfnamefont {J.}\ \bibnamefont
  {Preskill}},\ }\bibfield  {title} {\enquote {\bibinfo {title} {Quantum
  {C}omputing in the {NISQ} era and beyond},}\ }\href {\doibase
  10.22331/q-2018-08-06-79} {\bibfield  {journal} {\bibinfo  {journal}
  {{Quantum}}\ }\textbf {\bibinfo {volume} {2}},\ \bibinfo {pages} {79}
  (\bibinfo {year} {2018})}\BibitemShut {NoStop}%
\bibitem [{\citenamefont {Sieberer}\ and\ \citenamefont
  {Lechner}(2018)}]{Sieberer2018}%
  \BibitemOpen
  \bibfield  {author} {\bibinfo {author} {\bibfnamefont {L.~M.}\
  \bibnamefont {Sieberer}}\ and\ \bibinfo {author} {\bibfnamefont {W.}\
  \bibnamefont {Lechner}},\ }\bibfield  {title} {\enquote {\bibinfo {title}
  {Programmable superpositions of ising configurations},}\ }\href {\doibase
  10.1103/PhysRevA.97.052329} {\bibfield  {journal} {\bibinfo  {journal} {Phys.
  Rev. A}\ }\textbf {\bibinfo {volume} {97}},\ \bibinfo {pages} {052329}
  (\bibinfo {year} {2018})}\BibitemShut {NoStop}%
\bibitem [{\citenamefont {Hopfield}(1982)}]{Hopfield1982}%
  \BibitemOpen
  \bibfield  {author} {\bibinfo {author} {\bibfnamefont {J.~J.}\ \bibnamefont
  {Hopfield}},\ }\bibfield  {title} {\enquote {\bibinfo {title} {Neural
  networks and physical systems with emergent collective computational
  abilities},}\ }\href {http://www.pnas.org/content/79/8/2554.abstract}
  {\bibfield  {journal} {\bibinfo  {journal} {Proc. Natl. Acad. Sci. USA}\ }\textbf {\bibinfo {volume} {79}},\ \bibinfo {pages}
  {2554} (\bibinfo {year} {1982})}\BibitemShut {NoStop}%
\bibitem [{\citenamefont {Hebb}(1949)}]{Hebb1949}%
  \BibitemOpen
  \bibfield  {author} {\bibinfo {author} {\bibfnamefont {D.~O.}\ \bibnamefont
  {Hebb}},\ }\href@noop {} {\emph {\bibinfo {title} {The organization of
  behavior: A neuropsychological theory}}}\ (\bibinfo  {publisher} {Wiley, New York},\ \bibinfo {year} {1949})\BibitemShut {NoStop}%
\bibitem [{\citenamefont {Baldi}\ and\ \citenamefont
  {Venkatesh}(1987)}]{Baldi1987}%
  \BibitemOpen
  \bibfield  {author} {\bibinfo {author} {\bibfnamefont {P.}\ \bibnamefont
  {Baldi}}\ and\ \bibinfo {author} {\bibfnamefont {S.~S.}\ \bibnamefont
  {Venkatesh}},\ }\bibfield  {title} {\enquote {\bibinfo {title} {Number of
  stable points for spin-glasses and neural networks of higher orders},}\
  }\href {\doibase 10.1103/PhysRevLett.58.913} {\bibfield  {journal} {\bibinfo
  {journal} {Phys. Rev. Lett.}\ }\textbf {\bibinfo {volume} {58}},\ \bibinfo
  {pages} {913} (\bibinfo {year} {1987})}\BibitemShut {NoStop}%
\bibitem [{Note1()}]{Note1}%
  \BibitemOpen
  \bibinfo {note} {The states $\mathinner {|{x_n}\delimiter "526930B
  }=\mathinner {|{1011100\protect \dots }\delimiter "526930B }$ are regarded as
  product states in the Pauli $\sigma _z$ basis, with individual bits $x_{n,i}
  = 0,1$ corresponding to eigenvalues $\pm 1$ of $\sigma _z^{(i)}$}\BibitemShut
  {NoStop}%
\bibitem [{\citenamefont {Hopfield}\ \emph {et~al.}(1983)\citenamefont
  {Hopfield}, \citenamefont {Feinstein},\ and\ \citenamefont
  {Palmer}}]{Hopfield1983}%
  \BibitemOpen
  \bibfield  {author} {\bibinfo {author} {\bibfnamefont {J.~J.}\ \bibnamefont
  {Hopfield}}, \bibinfo {author} {\bibfnamefont {D.~I.}\ \bibnamefont
  {Feinstein}}, \ and\ \bibinfo {author} {\bibfnamefont {R.~G.}\ \bibnamefont
  {Palmer}},\ }\bibfield  {title} {\enquote {\bibinfo {title} {`unlearning' has
  a stabilizing effect in collective memories},}\ }\href
  {http://dx.doi.org/10.1038/304158a0} {\bibfield  {journal} {\bibinfo
  {journal} {Nature (London)}\ }\textbf {\bibinfo {volume} {304}},\ \bibinfo {pages}
  {158} (\bibinfo {year} {1983})}\BibitemShut {NoStop}%
\bibitem [{\citenamefont {Kleinfeld}\ and\ \citenamefont
  {Pendergraft}(1987)}]{Kleinfeld1987}%
  \BibitemOpen
  \bibfield  {author} {\bibinfo {author} {\bibfnamefont {D.}~\bibnamefont
  {Kleinfeld}}\ and\ \bibinfo {author} {\bibfnamefont {D.~B.}~\bibnamefont
  {Pendergraft}},\ }\bibfield  {title} {\enquote {\bibinfo {title}
  {{``Unlearning'' increases the storage capacity of content addressable
  memories}},}\ }\href
  {http://www.cell.com/biophysj/abstract/S0006-3495(87)83310-6} {\bibfield
  {journal} {\bibinfo  {journal} {Biophys. J.}\ }\textbf {\bibinfo {volume}
  {51}},\ \bibinfo {pages} {47} (\bibinfo {year} {1987})}\BibitemShut
  {NoStop}%
\bibitem [{\citenamefont {Fachechi}\ \emph {et~al.}(2018)\citenamefont
  {Fachechi}, \citenamefont {Agliari},\ and\ \citenamefont
  {Barra}}]{Fachechi2018}%
  \BibitemOpen
  \bibfield  {author} {\bibinfo {author} {\bibfnamefont {A.}\ \bibnamefont
  {Fachechi}}, \bibinfo {author} {\bibfnamefont {E.}\ \bibnamefont
  {Agliari}}, \ and\ \bibinfo {author} {\bibfnamefont {A.}\ \bibnamefont
  {Barra}},\ }\bibfield  {title} {\enquote {\bibinfo {title} {Dreaming neural
  networks: forgetting spurious memories and reinforcing pure ones},}\ }\href
  {https://www.sciencedirect.com/science/article/pii/S0893608019300176} {\bibfield
  {journal} {\bibinfo  {journal} {Neural Networks}\ }\textbf {\bibinfo {volume}
  {112}},\ \bibinfo {pages} {24} (\bibinfo {year} {2019})}\BibitemShut
  {NoStop}%
\bibitem [{\citenamefont {Lechner}\ \emph {et~al.}(2015)\citenamefont
  {Lechner}, \citenamefont {Hauke},\ and\ \citenamefont
  {Zoller}}]{Lechner2015}%
  \BibitemOpen
  \bibfield  {author} {\bibinfo {author} {\bibfnamefont {W.}\
  \bibnamefont {Lechner}}, \bibinfo {author} {\bibfnamefont {P.}\
  \bibnamefont {Hauke}}, \ and\ \bibinfo {author} {\bibfnamefont {P.}\
  \bibnamefont {Zoller}},\ }\bibfield  {title} {\enquote {\bibinfo {title} {{A
  quantum annealing architecture with all-to-all connectivity from local
  interactions}},}\ }\href
  {http://advances.sciencemag.org/content/1/9/e1500838.abstract} {\bibfield
  {journal} {\bibinfo  {journal} {Sci. Adv.}\ }\textbf {\bibinfo {volume}
  {1}},\ \bibinfo {pages} {e1500838} (\bibinfo {year} {2015})}\BibitemShut
  {NoStop}%
\bibitem [{\citenamefont {Abu-Mostafa}\ and\ \citenamefont
  {Jacques}(1985)}]{Mostafa1985}%
  \BibitemOpen
  \bibfield  {author} {\bibinfo {author} {\bibfnamefont {Y.}~\bibnamefont
  {Abu-Mostafa}}\ and\ \bibinfo {author} {\bibfnamefont {J.~St.}\ \bibnamefont
  {Jacques}},\ }\bibfield  {title} {\enquote {\bibinfo {title} {Information
  capacity of the hopfield model},}\ }\href {\doibase 10.1109/TIT.1985.1057069}
  {\bibfield  {journal} {\bibinfo  {journal} {IEEE Trans. Inf.
  Theory}\ }\textbf {\bibinfo {volume} {31}},\ \bibinfo {pages} {461}
  (\bibinfo {year} {1985})}\BibitemShut {NoStop}%
\bibitem [{\citenamefont {Baldi}(1988)}]{Baldi1988}%
  \BibitemOpen
  \bibfield  {author} {\bibinfo {author} {\bibfnamefont {P.}~\bibnamefont
  {Baldi}},\ }\bibfield  {title} {\enquote {\bibinfo {title} {Neural networks,
  orientations of the hypercube, and algebraic threshold functions},}\
  }\href {\doibase 10.1109/18.6032} {\bibfield
  {journal} {\bibinfo  {journal} {IEEE Trans. Inf. Theory}\
  }\textbf {\bibinfo {volume} {34}},\ \bibinfo {pages} {523} (\bibinfo
  {year} {1988})}\BibitemShut {NoStop}%
\bibitem [{\citenamefont {Hertz}\ \emph {et~al.}(1991)\citenamefont {Hertz},
  \citenamefont {Palmer},\ and\ \citenamefont {Krogh}}]{Hertz1991}%
  \BibitemOpen
  \bibfield  {author} {\bibinfo {author} {\bibfnamefont {J.}\ \bibnamefont
  {Hertz}}, \bibinfo {author} {\bibfnamefont {R.~G.}\ \bibnamefont
  {Palmer}}, \ and\ \bibinfo {author} {\bibfnamefont {A.~S.}\ \bibnamefont
  {Krogh}},\ }\href@noop {} {\emph {\bibinfo {title} {Introduction to the
  Theory of Neural Computation}}},\ \bibinfo {edition} {Santa Fe Institute Studies in the Sciences of Complexity, Lecture Notes, Redwood City, CA}\ (\bibinfo
  {publisher} {Addison-Wesley, Boston},\ \bibinfo {year} {1991})\BibitemShut
  {NoStop}%
\bibitem [{\citenamefont {Neigovzen}\ \emph {et~al.}(2009)\citenamefont
  {Neigovzen}, \citenamefont {Neves}, \citenamefont {Sollacher},\ and\
  \citenamefont {Glaser}}]{Neigovzen2009}%
  \BibitemOpen
  \bibfield  {author} {\bibinfo {author} {\bibfnamefont {R.}\ \bibnamefont
  {Neigovzen}}, \bibinfo {author} {\bibfnamefont {J.~L.}\ \bibnamefont
  {Neves}}, \bibinfo {author} {\bibfnamefont {R.}\ \bibnamefont
  {Sollacher}}, \ and\ \bibinfo {author} {\bibfnamefont {S.~J.}\
  \bibnamefont {Glaser}},\ }\bibfield  {title} {\enquote {\bibinfo {title}
  {Quantum pattern recognition with liquid-state nuclear magnetic resonance},}\
  }\href {\doibase 10.1103/PhysRevA.79.042321} {\bibfield  {journal} {\bibinfo
  {journal} {Phys. Rev. A}\ }\textbf {\bibinfo {volume} {79}},\ \bibinfo
  {pages} {042321} (\bibinfo {year} {2009})}\BibitemShut {NoStop}%
\bibitem [{\citenamefont {Santra}\ \emph {et~al.}(2017)\citenamefont {Santra},
  \citenamefont {Shehab},\ and\ \citenamefont {Balu}}]{Santra2017}%
  \BibitemOpen
  \bibfield  {author} {\bibinfo {author} {\bibfnamefont {S.}\
  \bibnamefont {Santra}}, \bibinfo {author} {\bibfnamefont {O.}\ \bibnamefont
  {Shehab}}, \ and\ \bibinfo {author} {\bibfnamefont {R.}\
  \bibnamefont {Balu}},\ }\bibfield  {title} {\enquote {\bibinfo {title} {Ising
  formulation of associative memory models and quantum annealing recall},}\
  }\href {\doibase 10.1103/PhysRevA.96.062330} {\bibfield  {journal} {\bibinfo
  {journal} {Phys. Rev. A}\ }\textbf {\bibinfo {volume} {96}},\ \bibinfo
  {pages} {062330} (\bibinfo {year} {2017})}\BibitemShut {NoStop}%
\bibitem [{\citenamefont {Fard}\ \emph {et~al.}(2018)\citenamefont {Fard},
  \citenamefont {Aghayar},\ and\ \citenamefont {Amniat-Talab}}]{Fard2018}%
  \BibitemOpen
  \bibfield  {author} {\bibinfo {author} {\bibfnamefont {E.~Rezaei}\
  \bibnamefont {Fard}}, \bibinfo {author} {\bibfnamefont {K.}~\bibnamefont
  {Aghayar}}, \ and\ \bibinfo {author} {\bibfnamefont {M.}~\bibnamefont
  {Amniat-Talab}},\ }\bibfield  {title} {\enquote {\bibinfo {title} {Quantum
  pattern recognition with multi-neuron interactions},}\ }\href {\doibase
  10.1007/s11128-018-1816-y} {\bibfield  {journal} {\bibinfo  {journal}
  {Quantum Inf. Process.}\ }\textbf {\bibinfo {volume} {17}},\
  \bibinfo {pages} {42} (\bibinfo {year} {2018})}\BibitemShut {NoStop}%
\bibitem [{\citenamefont {Rebentrost}\ \emph {et~al.}(2018)\citenamefont
  {Rebentrost}, \citenamefont {Bromley}, \citenamefont {Weedbrook},\ and\
  \citenamefont {Lloyd}}]{Rebentrost2018}%
  \BibitemOpen
  \bibfield  {author} {\bibinfo {author} {\bibfnamefont {P.}\ \bibnamefont
  {Rebentrost}}, \bibinfo {author} {\bibfnamefont {T.~R.}\ \bibnamefont
  {Bromley}}, \bibinfo {author} {\bibfnamefont {C.}\ \bibnamefont
  {Weedbrook}}, \ and\ \bibinfo {author} {\bibfnamefont {S.}\ \bibnamefont
  {Lloyd}},\ }\bibfield  {title} {\enquote {\bibinfo {title} {Quantum hopfield
  neural network},}\ }\href {\doibase 10.1103/PhysRevA.98.042308} {\bibfield
  {journal} {\bibinfo  {journal} {Phys. Rev. A}\ }\textbf {\bibinfo {volume}
  {98}},\ \bibinfo {pages} {042308} (\bibinfo {year} {2018})}\BibitemShut
  {NoStop}%
\bibitem [{\citenamefont {Seddiqi}\ and\ \citenamefont
  {Humble}(2014)}]{Seddiqi2018}%
  \BibitemOpen
  \bibfield  {author} {\bibinfo {author} {\bibfnamefont {H.}\ \bibnamefont
  {Seddiqi}}\ and\ \bibinfo {author} {\bibfnamefont {T.~S.}\ \bibnamefont
  {Humble}},\ }\bibfield  {title} {\enquote {\bibinfo {title} {Adiabatic
  quantum optimization for associative memory recall},}\ }\href {\doibase
  10.3389/fphy.2014.00079} {\bibfield  {journal} {\bibinfo  {journal}
  {Front. Phys.}\ }\textbf {\bibinfo {volume} {2}},\ \bibinfo {pages}
  {79} (\bibinfo {year} {2014})}\BibitemShut {NoStop}%
\bibitem [{\citenamefont {Rotondo}\ \emph {et~al.}(2018)\citenamefont
  {Rotondo}, \citenamefont {Marcuzzi}, \citenamefont {Garrahan}, \citenamefont
  {Lesanovsky},\ and\ \citenamefont {M{\"o}ller}}]{Rotondo2018}%
  \BibitemOpen
  \bibfield  {author} {\bibinfo {author} {\bibfnamefont {P.}~\bibnamefont
  {Rotondo}}, \bibinfo {author} {\bibfnamefont {M.}~\bibnamefont {Marcuzzi}},
  \bibinfo {author} {\bibfnamefont {J.~P.}\ \bibnamefont {Garrahan}}, \bibinfo
  {author} {\bibfnamefont {I.}~\bibnamefont {Lesanovsky}}, \ and\ \bibinfo
  {author} {\bibfnamefont {M.}~\bibnamefont {M{\"o}ller}},\ }\bibfield  {title}
  {\enquote {\bibinfo {title} {Open quantum generalisation of hopfield neural
  networks},}\ }\href {http://stacks.iop.org/1751-8121/51/i=11/a=115301}
  {\bibfield  {journal} {\bibinfo  {journal} {J. Phys. A:
  Math. Theor.}\ }\textbf {\bibinfo {volume} {51}},\ \bibinfo
  {pages} {115301} (\bibinfo {year} {2018})}\BibitemShut {NoStop}%
\bibitem [{\citenamefont {Amit}\ \emph
  {et~al.}(1985{\natexlab{a}})\citenamefont {Amit}, \citenamefont {Gutfreund},\
  and\ \citenamefont {Sompolinsky}}]{Amit1985PRL}%
  \BibitemOpen
  \bibfield  {author} {\bibinfo {author} {\bibfnamefont {D.~J.}\
  \bibnamefont {Amit}}, \bibinfo {author} {\bibfnamefont {H.}\ \bibnamefont
  {Gutfreund}}, \ and\ \bibinfo {author} {\bibfnamefont {H.}~\bibnamefont
  {Sompolinsky}},\ }\bibfield  {title} {\enquote {\bibinfo {title} {Storing
  infinite numbers of patterns in a spin-glass model of neural networks},}\
  }\href {\doibase 10.1103/PhysRevLett.55.1530} {\bibfield  {journal} {\bibinfo
   {journal} {Phys. Rev. Lett.}\ }\textbf {\bibinfo {volume} {55}},\ \bibinfo
  {pages} {1530} (\bibinfo {year} {1985}{\natexlab{a}})}\BibitemShut
  {NoStop}%
\bibitem [{\citenamefont {Amit}\ \emph
  {et~al.}(1985{\natexlab{b}})\citenamefont {Amit}, \citenamefont {Gutfreund},\
  and\ \citenamefont {Sompolinsky}}]{Amit1985PRA}%
  \BibitemOpen
  \bibfield  {author} {\bibinfo {author} {\bibfnamefont {D.~J.}\
  \bibnamefont {Amit}}, \bibinfo {author} {\bibfnamefont {H.}\ \bibnamefont
  {Gutfreund}}, \ and\ \bibinfo {author} {\bibfnamefont {H.}~\bibnamefont
  {Sompolinsky}},\ }\bibfield  {title} {\enquote {\bibinfo {title} {Spin-glass
  models of neural networks},}\ }\href {\doibase 10.1103/PhysRevA.32.1007}
  {\bibfield  {journal} {\bibinfo  {journal} {Phys. Rev. A}\ }\textbf {\bibinfo
  {volume} {32}},\ \bibinfo {pages} {1007} (\bibinfo {year}
  {1985}{\natexlab{b}})}\BibitemShut {NoStop}%
\bibitem [{\citenamefont {Rocchetto}\ \emph {et~al.}(2016)\citenamefont
  {Rocchetto}, \citenamefont {Benjamin},\ and\ \citenamefont
  {Li}}]{Rocchetto2016}%
  \BibitemOpen
  \bibfield  {author} {\bibinfo {author} {\bibfnamefont {A.}\ \bibnamefont
  {Rocchetto}}, \bibinfo {author} {\bibfnamefont {S.~C.}\ \bibnamefont
  {Benjamin}}, \ and\ \bibinfo {author} {\bibfnamefont {Y.}\ \bibnamefont
  {Li}},\ }\bibfield  {title} {\enquote {\bibinfo {title} {Stabilizers as a
  design tool for new forms of the {L}echner-{H}auke-{Z}oller annealer},}\
  }\href {\doibase 10.1126/sciadv.1601246} {\bibfield  {journal} {\bibinfo
  {journal} {Sci. Adv.}\ }\textbf {\bibinfo {volume} {2}},\ \bibinfo {pages} {e1601246}  (\bibinfo
  {year} {2016})}\BibitemShut {NoStop}%
\bibitem [{\citenamefont {Glaetzle}\ \emph {et~al.}(2017)\citenamefont
  {Glaetzle}, \citenamefont {van Bijnen}, \citenamefont {Zoller},\ and\
  \citenamefont {Lechner}}]{Glaetzle2017}%
  \BibitemOpen
  \bibfield  {author} {\bibinfo {author} {\bibfnamefont {A.~W.}\ \bibnamefont
  {Glaetzle}}, \bibinfo {author} {\bibfnamefont {R.~M.~W.}\ \bibnamefont {van
  Bijnen}}, \bibinfo {author} {\bibfnamefont {P.}~\bibnamefont {Zoller}}, \
  and\ \bibinfo {author} {\bibfnamefont {W.}~\bibnamefont {Lechner}},\
  }\bibfield  {title} {\enquote {\bibinfo {title} {A coherent quantum annealer
  with rydberg atoms},}\ }\href {http://dx.doi.org/10.1038/ncomms15813}
  {\bibfield  {journal} {\bibinfo  {journal} {Nat. Commun.}\ }\textbf
  {\bibinfo {volume} {8}},\ \bibinfo {pages} {15813} (\bibinfo {year}
  {2017})}\BibitemShut {NoStop}%
\bibitem [{\citenamefont {Leib}\ \emph {et~al.}(2016)\citenamefont {Leib},
  \citenamefont {Zoller},\ and\ \citenamefont {Lechner}}]{Leib2016}%
  \BibitemOpen
  \bibfield  {author} {\bibinfo {author} {\bibfnamefont {M.}\ \bibnamefont
  {Leib}}, \bibinfo {author} {\bibfnamefont {P.}\ \bibnamefont {Zoller}}, \
  and\ \bibinfo {author} {\bibfnamefont {W.}\ \bibnamefont {Lechner}},\
  }\bibfield  {title} {\enquote {\bibinfo {title} {A transmon quantum annealer:
  decomposing many-body ising constraints into pair interactions},}\ }\href
  {http://stacks.iop.org/2058-9565/1/i=1/a=015008} {\bibfield  {journal}
  {\bibinfo  {journal} {Quantum Sci. Technol.}\ }\textbf {\bibinfo
  {volume} {1}},\ \bibinfo {pages} {015008} (\bibinfo {year}
  {2016})}\BibitemShut {NoStop}%
\bibitem [{\citenamefont {Puri}\ \emph {et~al.}(2017)\citenamefont {Puri},
  \citenamefont {Andersen}, \citenamefont {Grimsmo},\ and\ \citenamefont
  {Blais}}]{Puri2017}%
  \BibitemOpen
  \bibfield  {author} {\bibinfo {author} {\bibfnamefont {S.}\ \bibnamefont
  {Puri}}, \bibinfo {author} {\bibfnamefont {C.~K.}\ \bibnamefont
  {Andersen}}, \bibinfo {author} {\bibfnamefont {A.~L.}\ \bibnamefont
  {Grimsmo}}, \ and\ \bibinfo {author} {\bibfnamefont {A.}\ \bibnamefont
  {Blais}},\ }\bibfield  {title} {\enquote {\bibinfo {title} {Quantum annealing
  with all-to-all connected nonlinear oscillators},}\ }\href
  {https://doi.org/10.1038/ncomms15785} {\bibfield  {journal} {\bibinfo
  {journal} {Nat. Commun.}\ }\textbf {\bibinfo {volume} {8}},\
  \bibinfo {pages} {15785} (\bibinfo {year} {2017})}\BibitemShut
  {NoStop}%
\bibitem [{\citenamefont {Chancellor}\ \emph {et~al.}(2017)\citenamefont
  {Chancellor}, \citenamefont {Zohren},\ and\ \citenamefont
  {Warburton}}]{Chancellor2017}%
  \BibitemOpen
  \bibfield  {author} {\bibinfo {author} {\bibfnamefont {N.}~\bibnamefont
  {Chancellor}}, \bibinfo {author} {\bibfnamefont {S.}~\bibnamefont {Zohren}},
  \ and\ \bibinfo {author} {\bibfnamefont {P.~A.}\ \bibnamefont {Warburton}},\
  }\bibfield  {title} {\enquote {\bibinfo {title} {Circuit design for
  multi-body interactions in superconducting quantum annealing systems with
  applications to a scalable architecture},}\ }\href {\doibase
  10.1038/s41534-017-0022-6} {\bibfield  {journal} {\bibinfo  {journal} {npj
  Quantum Information}\ }\textbf {\bibinfo {volume} {3}},\ \bibinfo {pages}
  {21} (\bibinfo {year} {2017})}\BibitemShut {NoStop}%
\bibitem [{\citenamefont {Albash}\ and\ \citenamefont
  {Lidar}(2018)}]{Albash2018}%
  \BibitemOpen
  \bibfield  {author} {\bibinfo {author} {\bibfnamefont {T.}\ \bibnamefont
  {Albash}}\ and\ \bibinfo {author} {\bibfnamefont {D.~A.}\ \bibnamefont
  {Lidar}},\ }\bibfield  {title} {\enquote {\bibinfo {title} {Adiabatic quantum
  computation},}\ }\href {\doibase 10.1103/RevModPhys.90.015002} {\bibfield
  {journal} {\bibinfo  {journal} {Rev. Mod. Phys.}\ }\textbf {\bibinfo {volume}
  {90}},\ \bibinfo {pages} {015002} (\bibinfo {year} {2018})}\BibitemShut
  {NoStop}%
\bibitem [{\citenamefont {Mandr\`a}\ \emph {et~al.}(2017)\citenamefont
  {Mandr\`a}, \citenamefont {Zhu},\ and\ \citenamefont
  {Katzgraber}}]{Katzgraber2017}%
  \BibitemOpen
  \bibfield  {author} {\bibinfo {author} {\bibfnamefont {S.}\
  \bibnamefont {Mandr\`a}}, \bibinfo {author} {\bibfnamefont {Z.}\
  \bibnamefont {Zhu}}, \ and\ \bibinfo {author} {\bibfnamefont {H.~G.}\
  \bibnamefont {Katzgraber}},\ }\bibfield  {title} {\enquote {\bibinfo {title}
  {Exponentially biased ground-state sampling of quantum annealing machines
  with transverse-field driving hamiltonians},}\ }\href {\doibase
  10.1103/PhysRevLett.118.070502} {\bibfield  {journal} {\bibinfo  {journal}
  {Phys. Rev. Lett.}\ }\textbf {\bibinfo {volume} {118}},\ \bibinfo {pages}
  {070502} (\bibinfo {year} {2017})}\BibitemShut {NoStop}%
\bibitem [{\citenamefont {K{\"o}nz}\ \emph {et~al.}()\citenamefont {K{\"o}nz},
  \citenamefont {Mazzola}, \citenamefont {Ochoa}, \citenamefont {Katzgraber},\
  and\ \citenamefont {Troyer}}]{Koenz2018}%
  \BibitemOpen
  \bibfield  {author} {\bibinfo {author} {\bibfnamefont {M.~S.}\
  \bibnamefont {K{\"o}nz}}, \bibinfo {author} {\bibfnamefont {G.}\
  \bibnamefont {Mazzola}}, \bibinfo {author} {\bibfnamefont {A.~J.}\
  \bibnamefont {Ochoa}}, \bibinfo {author} {\bibfnamefont {H.~G.}\
  \bibnamefont {Katzgraber}}, \ and\ \bibinfo {author} {\bibfnamefont
  {M.}\ \bibnamefont {Troyer}},\ }\bibfield  {title} {\enquote {\bibinfo
  {title} {Uncertain fate of fair sampling in quantum annealing},}\ }\href
  {https://arxiv.org/abs/1806.06081} {\bibinfo {journal} {arXiv:1806.06081}}\BibitemShut {NoStop}%
\bibitem [{\citenamefont {{De Grandi}}\ and\ \citenamefont
  {Polkovnikov}(2010)}]{DeGrandi2010}%
  \BibitemOpen
  \bibfield  {author} {\bibinfo {author} {\bibfnamefont {C.}~\bibnamefont {{De
  Grandi}}}\ and\ \bibinfo {author} {\bibfnamefont {A.}~\bibnamefont
  {Polkovnikov}},\ }\bibfield  {title} {\enquote {\bibinfo {title} {{Quantum
  Quenching, Annealing and Computation}},}\ \ }(\bibinfo{publisher} {Springer
  Berlin/Heidelberg},\ \bibinfo
  {year} {2010}), pp. \bibinfo {pages}
  {75--114}\BibitemShut {NoStop}%
\bibitem [{\citenamefont {Amin}(2009)}]{Amin2009}%
  \BibitemOpen
  \bibfield  {author} {\bibinfo {author} {\bibfnamefont {M.~H.~S.}\
  \bibnamefont {Amin}},\ }\bibfield  {title} {\enquote {\bibinfo {title}
  {Consistency of the adiabatic theorem},}\ }\href {\doibase
  10.1103/PhysRevLett.102.220401} {\bibfield  {journal} {\bibinfo  {journal}
  {Phys. Rev. Lett.}\ }\textbf {\bibinfo {volume} {102}},\ \bibinfo {pages}
  {220401} (\bibinfo {year} {2009})}\BibitemShut {NoStop}%
\bibitem [{\citenamefont {Bravyi}\ \emph {et~al.}(2011)\citenamefont {Bravyi},
  \citenamefont {DiVincenzo},\ and\ \citenamefont {Loss}}]{Bravyi2011}%
  \BibitemOpen
  \bibfield  {author} {\bibinfo {author} {\bibfnamefont {S.}\ \bibnamefont
  {Bravyi}}, \bibinfo {author} {\bibfnamefont {D.~P.}\ \bibnamefont
  {DiVincenzo}}, \ and\ \bibinfo {author} {\bibfnamefont {D.}\ \bibnamefont
  {Loss}},\ }\bibfield  {title} {\enquote {\bibinfo {title} {{Schrieffer-Wolff
  transformation for quantum many-body systems}},}\ }\href {\doibase
  10.1016/j.aop.2011.06.004} {\bibfield  {journal} {\bibinfo  {journal} {Ann.
  Phys. (NY)}\ }\textbf {\bibinfo {volume} {326}},\ \bibinfo {pages}
  {2793} (\bibinfo {year} {2011})}\BibitemShut {NoStop}%
\end{thebibliography}
\end{document}